\pdfoutput=1
\documentclass{JINST}
\usepackage{graphicx}
\usepackage{amsmath,amssymb}   
\usepackage{subfigure}
\usepackage{wasysym}
\usepackage{lineno}
\usepackage{cite}
\usepackage{bigints}
\usepackage{caption}
\usepackage{bm}
\usepackage{url}
\RequirePackage{mathptmx}      
\RequirePackage{flushend}
\RequirePackage[percent]{overpic}
\RequirePackage{setspace}

\usepackage{caption}
\usepackage{bm}
\title{On methods for correcting for the look-elsewhere effect in searches for new physics}

\author{S.~Algeri$^{1,2}$\thanks{Corresponding author.}, D.A.~van Dyk$^1$, J.~Conrad$^{1,2}$\thanks{Wallenberg Academy Fellow.}, B.~Anderson$^2$ \\
\llap{}$^1$Statistics Section, Department of Mathematics, Imperial College London, South Kensington Campus, London SW7 2AZ, United Kingdom\\
$^2$The Oskar Klein Centre for Cosmoparticle Physics, AlbaNova, SE-106 91 Stockholm, Sweden\\
\\
  E-mail: \email{s.algeri14@imperial.ac.uk}}

\abstract{The search for new significant peaks over a energy spectrum often involves a statistical multiple hypothesis testing problem. Separate tests of hypothesis are conducted at different locations  over a fine grid producing an ensemble of local p-values,  the smallest of which is reported as evidence for the new resonance. Unfortunately, controlling  the false detection rate (type I error rate) of such procedures may lead to excessively stringent acceptance criteria. In the recent physics literature, two promising statistical tools have been proposed to overcome these limitations. In 2005, a method to ``find needles in haystacks"     was introduced by Pilla et al. \cite{PL05}, and a second method was later proposed by Gross and Vitells \cite{gv10} in the context of the \textquotedblleft  look-elsewhere effect"    and trial factors.  We show that, although the two methods exhibit similar performance for large sample sizes, for relatively small sample sizes, the method of Pilla et al. leads to an artificial inflation of statistical power  that stems from an increase in the false detection rate. This method, on the other hand, becomes particularly useful in multidimensional searches, where the Monte Carlo simulations required by Gross and Vitells are often unfeasible. We apply the methods to realistic simulations of the Fermi Large Area Telescope data,
 in particular the search for dark matter annihilation lines. Further, we discuss the counter-intuitive scenario where the look-elsewhere corrections are more conservative than much more computationally efficient corrections for multiple hypothesis testing.
Finally, we provide general guidelines for navigating the tradeoffs between statistical and computational efficiency when selecting a statistical procedure for signal detection.}


\begin{document}

\section{Introduction}
\label{intro}
In High Energy Physics (HEP) the statistical evidence for new physics is determined using p-values, i.e., the probability of  observing a signal as strong or stronger than the one observed if the proposed new physics does not exist. If the location of the resonance in question is known, the p-value can be easily obtained with classical methods such as the Likelihood Ratio Test (LRT), using the asymptotic distribution provided under the conditions specified in Wilks or Chernoff's theorems \cite{wilks, chernoff}.
Unfortunately, the most realistic scenario involves signals with unknown locations, leading to what is known in the statistics literature as a non-identifiability problem \cite{davison}.

To tackle this difficulty, physicists traditionally considered multiple hypothesis testing: they scan the energy spectrum\footnote{The search of a new source emission  can occur over the spectrum of the mass, energy or any other physical characteristic; for simplicity, we will refer to it as energy spectrum.} 
over a predetermined number of locations (or grid points), and sequentially test for resonance in each location \cite{dellanegra, dvd14}.  As discussed in  detail in Section~\ref{MC}, when the number of grid points is large, the detection threshold for the resulting \emph{local} p-values becomes more anti-conservative than the  overall significance, which translates into a higher number of false discoveries than expected.  This is typically the case when the discretization of the search range is chosen fine enough to approximate the continuum of the energy window considered. We discuss the details of this phenomenon in Sections~\ref{tests} and~\ref{MC}. 

The situation is particularly problematic in the more realistic case of correlated tests. For instance, if the signal is dispersed over a wide energy range, its detection in a particular location may be correlated with that in nearby grid points. Unlike the case of uncorrelated tests in which the local significances can be determined exaclty, in presence of correlation,  we can only determine upper bounds for these significances, such as those provided by the Bonferroni's correction. Unfortunately, such bounds may often be excessively conservative \cite{MCref, conradDM}. We focus on the problem of finding a  single, or few peaks above background rather than multiple signals, and thus appealing methods such as Tukey's multiple comparisons \cite{tukey} or the popular False Discovery Rate (FDR) \cite{benjamini, efron, deep} do not apply in this scenario. 

In order to overcome some of the limitations arising in multiple hypothesis testing, two promising methods have been recently proposed in physics literature. The first (henceforth PL) was introduced in 2005 \cite{PL05} and refined in \cite{PL06}. Its methodology relies on the Score function and is purported to be more powerful than the usual  Likelihood Ratio Test (LRT) approach. Unfortunately, the mathematical implementation of the method is not straightforward, which strongly limited its diffusion within the physics community.  This  is one of the main motivations of this work. Specifically one of the questions we aim to address is if, despite its technical difficulties, PL provides some advantages in practical applications. It turns out that PL is particularly helpful for multi-dimensional signal searches. The second approach (hereinafter GV) belongs to the class of LRT-based methods. It was first introduced in 2010 \cite{gv10}, and recently extended  \cite{us15} to compare non-nested models. In contrast to PL, GV enjoys easy implementation, which has led to a wide range of applications in various searches for new physics including in the discovery of the Higgs boson \cite{dellanegra, dvd14, BEH1, BEH2}.  From a theoretical perspective, both approaches require an approximation of tail probabilities of the form $P(\sup{Y_t}>c)$, where $Y_t$ is either a $\chi^2$ or a Gaussian process. These approximations compute the distribution of the relevant test statistic evaluated at each possible signal location in the large-sample limit. GV formalizes the problem in terms of the number of times the process $Y_t$, when viewed as a function of the signal location, passes upward through the threshold $c$;  this is called the number of ``upcrossings''. PL, on the other hand, involves the so-called tube formulae, where an approximation of $P(\sup{Y_t}>c)$ is obtained as the ratio between the volume of a tube  built around the manifold associated with $\sup{Y_t}$ on the unit sphere, and the volume of the unit sphere itself. Although we describe both methods more fully in Section \ref{methods}, we do not focus on their mathematical details, but rather emphasize their computational implementation; readers are directed to \cite{davies77, davies87, gv10, us15, PL05,PL06, adler2000} for technical development.

While either GV or PL can be used to control the false detection rate and ensure sufficient statistical power, they can be computationally expensive in complex models.  GV specifically, may easily become unfeasible in the multidimensional scenario. Multiple hypothesis testing procedures, on the other hand, can be much quicker, but are often overly conservative in terms of the false detection rate when the number of tests is large. Perhaps counter-intuitively, however, situations do occur where multiple hypothesis testing lead to the same or even less conservative inference than GV and PL. Not surprisingly, this depends on the number of tests conducted, i.e., GV and PL bounds on  p-values are less likely to be
larger than the Bonferroni's bound as the number of tests increases. In the absence of specific guidelines 
as to the optimal number of tests to conduct, and  in order to optimize computational speed while adhering to a prescribed false-positive rate as closely as possible, we summarize our findings as a simple algorithm that implements a sequential selection of the statistical procedure. Although it is well known that choosing a statistical procedure on the basis of its outcome can detrimentally effect the statistical significance, an effect called ``flip-flopping'' by Feldman and Cousins \cite{flipflop}, we show that our \emph{sequential procedure} is immune to this effect.

The remainder of this paper is organized as follows: in Section~\ref{tests} we review the background of hypothesis testing, we define the auxiliary concepts of \emph{goodness} of a test and \emph{local power}, which are used for our comparison of PL and GV. In Section~\ref{MC}, we review the multiple hypothesis testing approach for signal detection and we underline the respective disadvantages in terms of significance requirements. In Section~\ref{methods}, we provide a simplified overview of the technical results of PL and GV. In Section~\ref{simulation}, a suite of simulation studies is used to highlight the performance of the two methods in terms of approximation to the tail probabilities, false detection rate and statistical power. We show that both solutions exhibit advantages and suffer limitations, not only in terms of computational requirements and statistical power, but most importantly, in terms of the specific conditions they require of the models being tested.   An application to a realistic data simulation  is conducted in Section~\ref{application}. The sequential approach is discussed in Section~\ref{step} and discussion in Section~\ref{discussion}.

\section{Type I error,  local power and good tests of hypothesis}
\label{tests}
Consider the framework of a classical detection problem. Suppose $N$ event counts are observed over a predetermined energy band $\mathcal{Y}$. We are interested in knowing if some of  these events are due to a new emission source or if they all can be attributable to the background and its random fluctuations. We further assume that if there is no new  source, the energy $y$ of the $N$ events can be modeled using a probability density function (pdf) $f(y,\mathbf{\phi})$  over $\mathcal{Y}$ where $\mathbf{\phi}$ is a potentially unknown free parameter. Whereas, if the new resonance is present, events associated with it have energy distribution $g(y,\theta)$ over $\mathcal{Y}$, and we let $\theta \in \Theta$ with $\Theta$ representing the search window for the new resonance over the energy range. Typically $\Theta \equiv  \mathcal{Y}$, but in principle one could consider $\Theta \subset  \mathcal{Y}$.  Thus, we can write the full model for $N$ counts as
\begin{equation}
\label{add}
(1-\eta)f(y,\mathbf{\phi})+\eta g(y,\theta),
\end{equation} 
where $\eta$ is the source strengh, and positive values of $\eta$ indicate the presence of the new signal. 

From a statistical perspective, the search for new physics corresponds to a test of hypothesis in which the \emph{null hypothesis}, $H_0$, which stipulates that only background counts are observed, is tested against the \emph{alternative hypothesis}, $H_1$, which stipulates a proportion $\eta$ of the observed counts are due to new physics. Notationally this test is written 
\begin{equation}
\label{test}
H_0: \eta=0 \quad \mbox{versus} \quad H_1:\eta>0.
\end{equation}
The test is then conducted by specifying an opportune test statistic $T$, whose observed value $t_{obs}$ is calculated on the available data, and a detection is claimed if  $t_{obs}$ exceeds a specified detection threshold $t_{\alpha}$. The latter is  determined by controlling the probability of a \emph{type I error} or the false detection rate, which we allow to be no larger than a predetermined level $\alpha$. For obvious reasons, it is sensible to choose $\alpha$ sufficiently small, and it is common practice in physics to adopt a 3, 4 or 5$\sigma$ thresholds i.e., 
\begin{equation}
\label{phi}
\alpha=1-\Phi(x) \qquad x=3, 4, 5,
\end{equation}
where $\Phi(\cdot)$ is the cumulative density function (cdf) of a standard normal distribution. 
If $t_{obs}>t_{\alpha}$ a discovery is claimed, whereas if  $t_{obs}\leq t_{\alpha}$ we conclude that there is no sufficient evidence to claim detection of a new signal. 

An equivalent formulation of a test of hypothesis can be made in terms of a \emph{p-value} i.e., the probability of observing a value of $T$ that, under the hypothesis of no signal emission ($H_0$), is greater than $t_{obs}$. Formally
\begin{equation}
\label{pvalue}
\text{p-value}=P(T\geq t_{obs}|\eta=0).
\end{equation}
The p-value is then compared to the target probability of a type I error, $\alpha$. In this case, a discovery is claimed if $\text{p-value}<\alpha$, whereas the new resonance is not detected if  $\text{p-value}\geq \alpha$.

In addition to the type I error, another important property of a test of hypothesis is its statistical \emph{power} i.e., the probability of detecting the new signal when it is present. For the test in (\ref{test}) we can write
\begin{equation}
\label{alphapower}
\begin{split}
\alpha & = P(T>t_{\alpha}|\eta=0)\\
\text{Power$(\eta,\theta)$} & = P(T>t_{\alpha}|\eta, \theta), \quad  \eta>0.\\
\end{split}
\end{equation}
The goal is to construct a \emph{good} detection test, that is, a test with the probability of false detection, equal to or smaller than the predetermined level $\alpha$, but with the power  as large as possible. 

Consequently, if two or more tests with the same level $\alpha$ are to be compared, the test with higher power is preferred. As specified in \eqref{alphapower}, for the model in (\ref{add}) the power depends on  both the signal strength $\eta$ and its location $\theta$. For $\eta$, the detection power can be summarized using \emph{upper limits} as discussed in \cite{vinay}, whereas in this paper, we focus on the power with respect to the source location. This  is of particular importance when the dispersion of the signal depends on its position (as in our examples in Section~\ref{simulation}), and  widely spread source signals are expected to be more difficult to detect, i.e., exhibit lower statistical power.  Hereafter, we  refer to the power at a fixed location $\theta$ as the  \emph{local power}, and we say that a test is \emph{uniformly more powerful locally} than another test with the same level $\alpha$, if, for fixed $\eta$, its local power is greater than or equal to that  of the other test, for every possible $\theta$ in the energy range $\Theta$.
We  investigate the \emph{goodness} and the local power of PL and GV in Section~\ref{methods}. 

Typically, the exact distribution of the test statistic $T$ cannot be specified explicitely, and classical statistical methods  rely on its asymptotic distribution. It follows that the resulting p-values, $\alpha$, and power are also asymptotic quantities. In this paper, we mainly consider the asymptotic distributions of  various test statistics and thus, the p-values, $\alpha$ levels and powers that we quote are implicitly asymptotic quantities.   The only exceptions are the values quoted in the simulation studies in Section~\ref{simulation}.  There, the distribution of reference is the simulated distribution of $T$, and we  refer to the quantities of interest as simulated false detection rate and simulated power.

\section{Signal detection via multiple hypothesis testing}
\label{MC}
As anticipated in Section~\ref{intro}, the statistical detection of new physics can often be viewed as a multiple hypothesis testing problem. An ensemble of $R$ tests are conducted simultaneously, any of which can result in a false detection. While the individual tests are designed to control their specific false detection rate, the overall probability of having at least one false detection increases as $R$ increases, leading to a higher rate of false discoveries than expected. 

For the test in (\ref{test}), a natural choice of the test statistic $T$ is the LRT. Define 
\begin{equation}
\label{LRT}
LRT_{\theta}=-2\log\frac{L(0,\hat{\mathbf{\phi}}_0,\text{-})}{L(\hat{\eta}_1,\hat{\mathbf{\phi}}_1,\theta)},
\end{equation}
where $L(\eta,\mathbf{\phi},\theta)$ is the likelihood function 
under (\ref{add}). Notice that under $H_0$ (i.e., $\eta=0$), the parameter $\theta$ has no meaning and no value. The numerator and denominator of (\ref{LRT}) are the maximum likelihood achievable under $H_0$ and $H_1$ respectively, with $\hat{\mathbf{\phi}}_0$ being the Maximum Likelihood Estimate (MLE) of $\mathbf{\phi}$ under $H_0$ and $\hat{\mathbf{\phi}}_1$ and $\hat{\eta}_1$ the MLEs under $H_1$.  Under $H_0$, the distribution of the data does not depend on $\theta$. Because this violates a key assumption of both Wilks or Chernoff's theorems \cite{wilks, chernoff}, the distribution of LRT is not known and  we cannot directly compute the p-value for (\ref{test}). 

 To overcome this difficulty, a na\"ive approach involves the discretization of the energy range $\Theta$  into $R$ search regions,  resulting in a grid of fixed values $\Theta_\mathrm{G}=\{\theta_1,\dots,\theta_R\}$.
$R$ simultaneous LRTs are then conducted for the hypotheses in (\ref{test}), fixing $\theta$ in (\ref{LRT}) to be equal to each of the $\theta_r\in \Theta_\mathrm{G}$. In this way, a set of $R$ \emph{local} p-values is produced, and the smallest, namely $p_{L}$, is  compared with the established target probability of type I error, $\alpha_\mathrm L$. Notice that $\alpha_\mathrm L$ corresponds to the  false detection rate for a specific test among the $R$ available, and thus is the local significance. However, we must take account of the fact that  $R$ hypotheses are being tested simultaneously and  must also consider the chance of having at least one false detection among the ensemble of  $R$ tests, namely the global significance, $\alpha_\mathrm G$. 

If the $R$ tests are independent, i.e., detecting a signal in a given energy location does not depend on its detection  in other locations,  it can be easily shown  \cite{MCref} that
\begin{equation}    
\label{independent}
\alpha_\mathrm G=1-(1-\alpha_\mathrm L)^{R},
\end{equation}
and the resulting adjusted (global) p-value \cite{MCref,conradDM} is 
\begin{equation}    
\label{independentp}
p_\mathrm G=1-(1-p_\mathrm L)^{R}.
\end{equation}
Consider a toy example in which we have, 50  grid points over the energy spectrum $\mathcal{Y}$ and  50 uncorrelated tests at the $5\sigma$ significance level, the chance of having at least one false detection among  the 50 tests, i.e., the overall   false detection rate, is $\alpha_\mathrm G=1.4\cdot 10^{-5}$ which corresponds to $4.18\sigma$ significance. This is approximately 50 times larger than the $\alpha_\mathrm L=2.87\cdot10^{-7}$ associated with $5\sigma$.
Conversely, if the $R$ tests are correlated, as in the case of disperse source emission, controlling for the  false detection rate is more problematic. In this scenario, contrary to (\ref{independent}), an exact general relationship between $\alpha_\mathrm L$ and $\alpha_\mathrm G$ cannot be established, since the specific correlation structure varies on a case-by-case basis. Thus, the only general statement that we can make is
\begin{equation}    
\label{dependent}
\alpha_\mathrm G\leq R\alpha_\mathrm L.
\end{equation}
The adjusted  p-value corresponding to (\ref{dependent}) is known as the Bonferroni correction \cite{MCref}, specifically, 
\begin{equation}    
\label{bonferroni}
p_\mathrm{BF}=Rp_\mathrm{L}
\end{equation}
which bounds $p_\mathrm G$ in that $p_\mathrm G\leq p_\mathrm{BF}$. In particular, $p_\mathrm{BF}$ is a first order approximation of  $p_\mathrm G$, and  thus the two p-values are equivalent when dealing with strong signals, i.e., when $p_\mathrm L \rightarrow 0$. This  is reflected in the toy example above, where $p_\mathrm{BF}$ is equal to $p_\mathrm G$, and also leads to $4.18\sigma$ significance. (Recall $\frac{\alpha_\mathrm G}{ \alpha_\mathrm L} \approx 50$ in the toy example.)

Despite their easy implementation, these procedures are often dismissed by practitioners  because, in addition to the stringent  requirements  to control for the overall false detection rate, they artificially depend on the number of tests $R$. This is particularly troublesome given the typically arbitrary nature of setting $R$ when discretizing the energy spectrum $\Theta$. We discuss below, however, practical situations in which these methods provide reasonable inference and  occasionally perform better than the often preferred look-elsewhere corrections of GV and PL.

\section{Needles in haystacks and look elsewhere effect}
\label{methods}
In this section we consider methods that directly address problems associated with parameters that are only present under $H_1$. Rather than constructing $R$ tests, these methods consider a single test of hypothesis 
and a single global p-value. The key element of these methods is to consider  new test statistics, which are not affected by the non-identifiability of the parameters. The two methods we consider follow a similar overall strategy which we now summarize. 

Consider the model  in (\ref{add}). We denote the MLE of the parameters  $\eta$ and $\mathbf{\phi}$ by $\hat{\mathbf{\phi}}_\theta, \hat{\eta}_\theta$ for each  fixed value $\theta \in \Theta$, and we specify a \emph{local} test statistic $C(y,\hat{\mathbf{\phi}}_\theta,\hat{\eta}_\theta,\theta)$ for the test in (\ref{test}). For brevity, we write  $C(y,\hat{\mathbf{\phi}}_\theta,\hat{\eta}_\theta,\theta)$ as $C(\theta)$. In practice, for each fixed value $\theta_r \in \Theta_\mathrm{G}$,  we compute $c(\theta_1),\dots,c(\theta_R)$, where $c(\theta_r)$ corresponds to the observed value of $C(\theta)$ with $\theta=\theta_r$. The collection of values $\{c(\theta_1),\dots,c(\theta_R)\}$ can be viewed as a realization of a stochastic process  $\{C(\theta), \theta \in \Theta \}$, and a \emph{global} test statistic, for (\ref{test}) is 
\begin{equation}
\label{supremum}
C=\sup_{\theta \in \Theta}C(\theta).
\end{equation}
Because we only observe $C(\theta)$ for $\theta_r \in \Theta_ \mathrm G$, the observed value of $C$ is
\begin{equation}
\label{cobs}
c(\hat{\theta})=\max_{\theta_r \in \Theta_\mathrm G} c(\theta_r)
\end{equation}
 where $\hat{\theta}$  is the value $\theta_r \in \Theta_\mathrm{G} $ where this maximum is attained, and which corresponds to  our estimate of the signal location. Finally, the \emph{global} p-value of the test is obtained by approximating the tail probability 
\begin{equation}
\label{tail}
P(C>c(\hat{\theta}))
\end{equation}
 under $H_0$. 
The choice of the statistic $C$ and the approximation method for computing (\ref{tail}) are the main characteristics differentiating the approaches of PL and GV.

To derive $C$, PL \cite{PL05,PL06} considers the Score process $\{C^{\star}_\mathrm{PL}(\theta),$ $\theta \in \Theta \}$, with 
\begin{equation}
\label{CPLstar}
C^{\star}_\mathrm{PL}(\theta)=\sum_{i=1}^N\biggl [\frac{f(y_i,\mathbf{\phi})}{g(y_i,\theta)}-1\biggl]
\end{equation}
being the Score function of (\ref{add}) under $H_0$ and  the generic local statistic $C(\theta)$ above is replaced by the normalized Score function, 
\begin{equation}
\label{Sstat}
C_\mathrm{PL}(\theta)=\frac{C^\star_\mathrm{PL}(\theta)}{\sqrt{NW(\theta,\theta)}}
\end{equation}
where $W(\theta,\theta^{\dag})$ is the covariance function  of $\{C^{\star}_\mathrm{PL}(\theta), \theta \in \Theta \}$. The functional form  of $W(\theta,\theta)$ depends on whether the free parameter under $H_0$, $\mathbf{\phi}$, is  known or not (see Appendix \ref{appendix1}).

The stochastic process of interest is $\{C_\mathrm{PL}(\theta), \theta \in \Theta \}$ and we let $C_\mathrm{PL}=\sup_{\theta \in \Theta} C_\mathrm{PL}(\theta)$ and $c_\mathrm{PL}(\hat{\theta})$ be its observed value. In order to simplify notation we drop the dependence of $c_\mathrm{PL}(\hat{\theta})$ on $\hat{\theta}$ and write simply, $c_\mathrm{PL}$. The corresponding global p-value  is $P(C_\mathrm{PL}>c_\mathrm{PL})$; \cite{PL06} prove that, under $H_0$, $C_\mathrm{PL}$
 converges to the supremum of a mean zero Gaussian process as $N\rightarrow \infty$. The approximation, $p_\mathrm{PL}$, of $P(C_\mathrm{PL}>c_\mathrm{PL})$ is  obtained through so-called tube formulae for Gaussian processes \cite{adler2000}. 
In particular, the supremum of the Gaussian (large-sample) limiting process of $\{C_\mathrm{PL}(\theta), \theta \in \Theta \}$  is approximated via an appropriate one-dimensional  manifold over a unit sphere; a tube is then constructed around the manifold and the ratio of the volume of the tube and of a unit sphere is used to approximate $P(C_\mathrm{PL}>c_\mathrm{PL})$.   If $\theta$ is
one-dimensional, the approximation to   $P(C_\mathrm{PL}>c_\mathrm{PL})$ is
\begin{equation}
\label{PLaprox}
p_\mathrm{PL} = \frac{\xi_0}{2\pi}P(\chi^2_{2}\geq c_\mathrm{PL}^2) + \frac{1}{2} P(\chi^2_{1}\geq c_\mathrm{PL}^2),
\end{equation}
which becomes more precise as $c_\mathrm{PL}\rightarrow \infty$, and where in general $P(\chi^2_{s}\geq q)=1-P(\chi^2_{s}< q)$, with $P(\chi^2_{s}< q)$ being the cumulative density distribution of a $\chi^2$ random variable with $s$ degrees of freedom evaluated at $q$. The quantity $\xi_0$ in \eqref{PLaprox} is the volume of the one-dimensional manifold (see Appendix \ref{appendix2} for more details). 
\begin{figure}
\begin{tabular*}{\textwidth}{@{\extracolsep{\fill}}@{}c@{}}
      \includegraphics{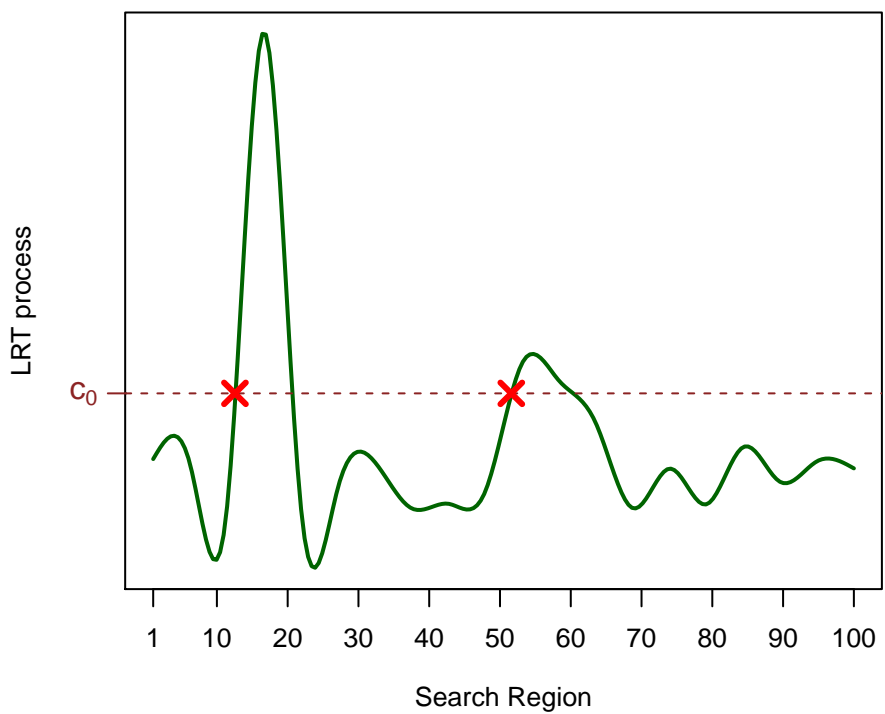}\\
\end{tabular*}
\caption{Upcrossings (red crosses) of the threshold $c_0$ by the LRT process. }
\label{upcrossings}
\end{figure}

Instead of the Score function, GV   \cite{gv10} focuses on the LRT in (\ref{LRT}), and thus $C_\mathrm{GV}(\theta)=LRT(\theta)$. For the specific case of (\ref{test}), $H_0$ is on the boundary of the parameter space, and thus under $H_0$ the LRT process converges asymptotically to a   $\frac{1}{2}\chi^2_1+\frac{1}{2}\delta(0)$ random process \cite{gv10,us15}. With this choice, and again, dropping the dependence on $\hat{\theta}$, we let $C_\mathrm{GV}=\sup_{\theta \in \Theta} C_\mathrm{GV}(\theta)$ and  $c_\mathrm{GV}$ be its observed value depending on the data. The global p-value $P(C_\mathrm{GV}>c_\mathrm{GV})$,  is approximated by 
\begin{equation}
\label{GVaprox}
p_\mathrm{GV}= \frac{P(\chi^2_{1}>c_\mathrm{GV})}{2} +E[U(c_0)|H_0]e^{-\frac{c_\mathrm{GV}-c_0}{2}}.
\end{equation}
which becomes more precise as  $c_\mathrm{GV} \rightarrow \infty$ and where $c_0$ is a small threshold such that  $c_0<<c_\mathrm{GV}$, and $U(c_0)$ is the number of times the LRT process, when viewed as a function of $\theta$, crosses from below $c_0$ to above $c_0$; this is called the number of upcrossings. An illustrative example is shown in Figure \ref{upcrossings}.  In \eqref{GVaprox}, $E[U(c_0)|H_0]$ is the expected number of upcrossings under $H_0$ of the (large-sample) LRT process, and is estimated via a Monte Carlo simulation of size $M$ as described in Algorithm~1.

\vspace{0.5cm}
\textbf{\underline{\emph{Algorithm 1.}}}
\begin{itemize}
\item For $m=1,\dots, M$:
\begin{enumerate}
\item[(1) - ] Simulate a large number (e.g., 1,000) of observations from $f(y,\hat{\mathbf \phi}_0)$;
\item[(2) - ] for each $\theta_r \in \Theta_G$ calculate $LRT(\theta_r)$ as in \eqref{LRT};
\item[(3) - ] for each $r \in [1;R-1]$ count how many times $LRT(\theta_{r})<c_0$ and $LRT(\theta_{r+1})\geq c_0$, i.e., the number  of upcrossings of $c_0$ by the LRT process under $H_0$ for simulation $m$, namely, $U_m(c_0)$.
\end{enumerate}
\item Estimate $E[U(c_0)|H_0]$ with $\frac{1}{M}\sum_{m=1}^MU_m(c_0)$.
\end{itemize}
\vspace{0.3cm}

The threshold $c_0$ is typically chosen to be small enough so that a reliable estimate of $E[U(c_0)|H_0]$ can be obtained with a small Monte Carlo simulation size $M$, but large enough so that the effect of the resolution $R$ of $\Theta_\mathrm{G}$ on the number of upcrossings is negligible (see \cite{gv10}). Although  (\ref{PLaprox}) and (\ref{GVaprox}) both hold when $c_\mathrm{PL}$ and $c_\mathrm{GV}$ are large, 
when they are small, the right hand sides of (\ref{PLaprox}) and (\ref{GVaprox}) provide  upper bounds for the respective tail probabilities.

GV's global p-value, $p_\mathrm {GV}$, is always greater than or equal to the smallest local p-value, $p_\mathrm L$, introduced in Section~\ref{MC}.  Thus GV always leads to an equal  or smaller number of false discoveries than one would have using multiple hypothesis testing when no correction is applied. This can be easily shown by noticing that for the test in (\ref{test})
\begin{equation}
\label{pL}
p_\mathrm L=\frac{1}{2} P(\chi^2_1>LRT_{\theta^\star})
\end{equation}
where $LRT_{\theta^\star}$ is calculated according to (\ref{LRT}) with $\theta= \theta^\star$. Notice that  $\theta^\star\equiv \hat{\theta}$, i.e., the location where the smallest p-value is observed is also where the  observed local LRT statistic, achieves its maximum. Thus,  the $LRT_{\theta^\star}$ coincides with the observed value  $c_\mathrm{GV}$ of the GV test statistic $C_\mathrm{GV}$. It follows by (\ref{GVaprox}) and (\ref{pL}) that the inequality $p_\mathrm{GV} \geq  p_\mathrm L$ 
always holds.

\begin{figure*}
\begin{tabular*}{\textwidth}{@{\extracolsep{\fill}}@{}c@{}c@{}}
      \includegraphics[width=75mm]{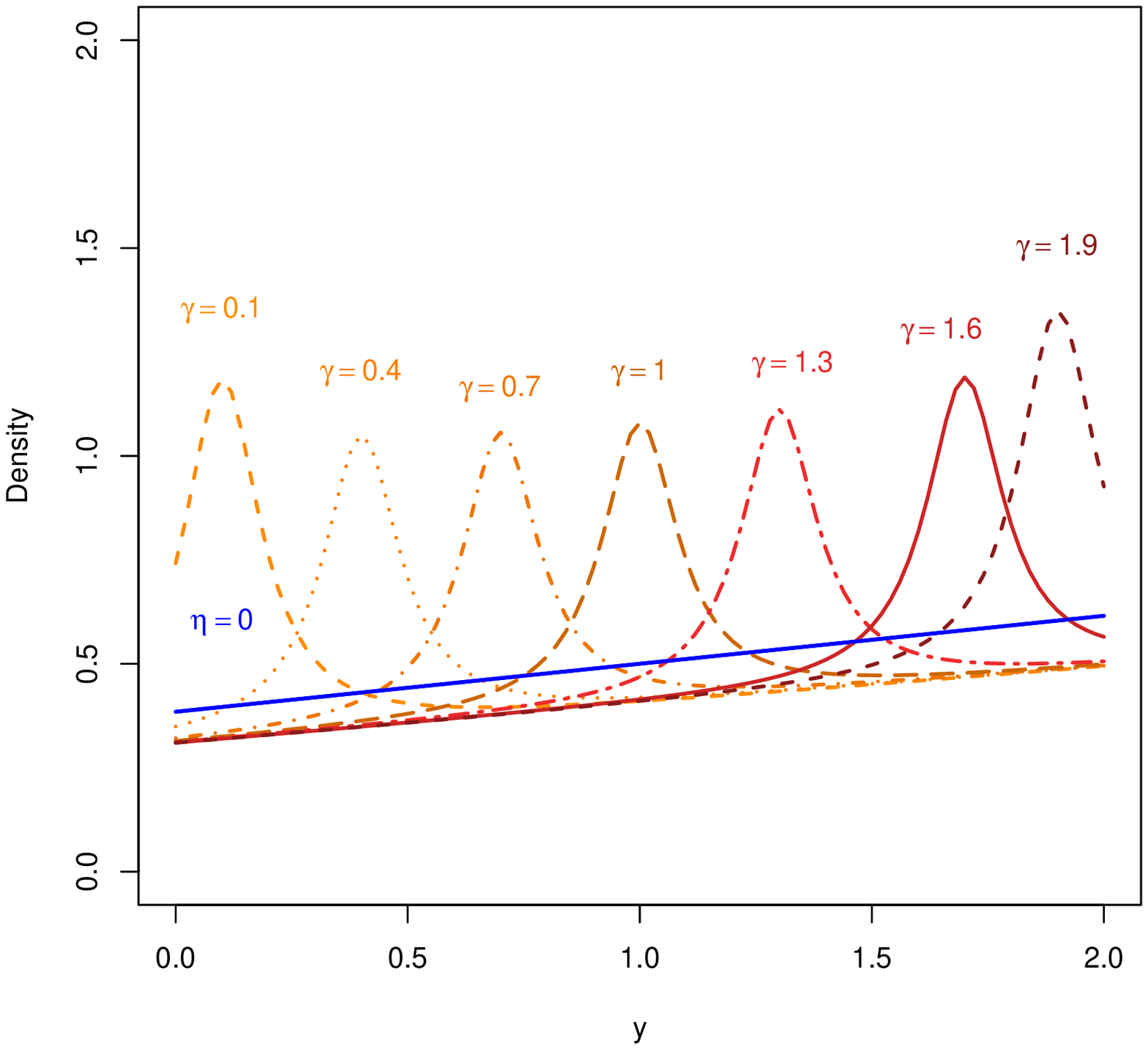} & \includegraphics[width=75mm]{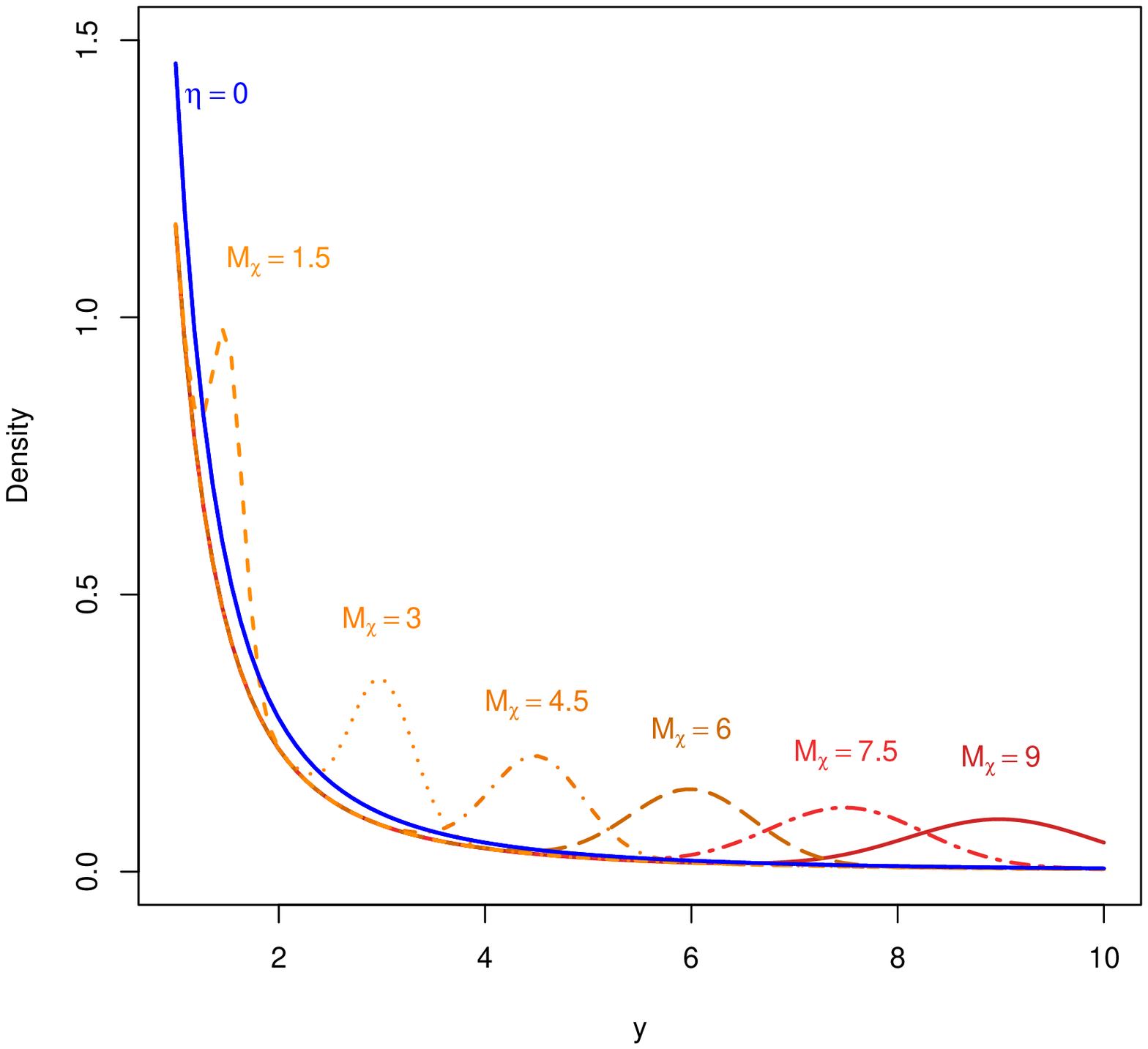}\\
\end{tabular*}
\caption[Figure 2]{Left panel: probability density functions for Example~I  under $H_0$ (blue line) and $H_1$ (orange lines) with $\eta=0.2$ and $\gamma=0.1, 0.4,0.7, 1, 1.3,1.6,1.9$. Right panel: probability density functions for Example~II  under $H_0$ (blue line) with $\tau=1.4$ and $H_1$ (orange lines) with $\eta=0.2$ and $M_{\chi}=1.5, 3, 4.5, 6, 7.5, 9$. }
\label{densities}
\end{figure*}

Another fundamental difference between the multiple hypothesis testing approach in Section~\ref{MC} and the  methods discussed in this section is the level at which the optimization occurs. In the former, the $p_\mathrm L$ is the minimum of set of local p-values  \[p_\mathrm L=\min_{\theta_r \in \Theta_\mathrm{G}}p(\theta_r),\]
and the result, is eventually corrected afterwards according to (\ref{independentp}) or (\ref{bonferroni}). Conversely, as expressed in (\ref{cobs}) in PL and GV, the optimization occurs with respect to the statistic $C(\theta)$,
and a correction for $p_\mathrm L$ is eventually generated intrinsically, by approximating the tail probability of the test statistic $C$.

\section{Simulation studies}
\label{simulation}
A fundamental result in probability theory states that the Score test and the LRT are asymptotically equivalent when the number of events is large (i.e., for large sample sizes). As shown in \cite{PL05}, the same can be proven for the $C_\mathrm{PL}$ and $C_\mathrm{GV}$ of PL and GV, respectively,  and thus, we expect the asymptotic equality between $p_{\mathrm L}$ and $p_{\mathrm GV}$ to hold for $p_{\mathrm PL}$, at least for large sample sizes.

Unfortunately, as one might expect, the asymptotic equivalence does not necessarily hold for small sample sizes, i.e., when only a few counts are available.
In order to  investigate this scenario, we consider two examples. In Example~I, we refer to  the toy model in \cite{PL05} where a Breit-Wigner resonance is superimposed on a linear background. The full model is
\begin{equation}
\label{ex1}
(1-\eta)\frac{1+0.3y}{2.6}+\eta \frac{0.1}{k_{\gamma}\pi(0.01+(y-\gamma)^2)}
\end{equation}
where $k_{\gamma}$ is a normalizing constant, $y\in[0;2]$ and  $\gamma\in(0;2]$. Notice that the null model has no free parameters and thus PL can be directly applied with no further adjustment of the covariance function (see Section~\ref{methods}). In Example~II, the background is  power-law distributed with unknown parameter $\tau$.  The signal component is modeled as a Gaussian bump  with dispersion proportional to the signal  location. Specifically, the full model is
\begin{equation}
\label{ex2}
(1-\eta)\frac{1}{k_{\tau}y^{\tau+1}}+ \frac{\eta }{k_{M_{\chi}}}\exp\biggl\{-\frac{(y-M_{\chi})^2}{0.02M^2_{\chi}}\biggl\}
\end{equation}
with $k_{\tau}$ and $k_{M_{\chi}}$  normalizing constants, $y\in[1;10]$, $\tau>0$ and $M_{\chi}\in[1;10]$. Owing to the unknown parameter $\tau$ under $H_0$, we must use the extended theory in \cite{PL06} for PL. 
 The  pdfs used in Example~I and II  are plotted in Fig.~\ref{densities}.

For both  examples, we evaluate the false detection rate (or type I error), and the local power as described in Section~\ref{tests}, and examine how it depends on the number of events; specifically, we considered sample sizes of $10, 50, 100, 200$ and $500$. The false detection rate and local power are obtained via Monte Carlo simulations from the null model ($\eta=0$) and from the alternative model with $\eta=0.2$, respectively. Although $\tau$ is unknown in Example~II, it can be estimated with the MLE $\hat{\tau}$ under $H_0$. The simulations are then drawn from (\ref{ex2}) with $\tau=\hat{\tau}$. This simulation procedure is known in the statistical literature as the parametric bootstrap \cite{boot}. In principle, the observed sample used to compute $\hat{\tau}$ could either come from the null  or from the alternative model. Thus, in order to evaluate the consistency of PL and GV in both situations, two further sub-cases are needed. In Example~IIa, we draw the ``observed" sample from (\ref{ex2}) with $\eta=0$ and $\tau=1.4$, i.e., in absence of new physics. In Example~IIb, we draw the ``observed" sample with $\eta=0.2$, $\tau=1.4$ and $M_{\chi}=9$.

\begin{figure*}
\noindent
\hspace*{-2cm}
\bgroup
\def\arraystretch{1}
\begin{tabular*}{\textwidth}{@{\extracolsep{\fill}}@{}c@{}c@{}c@{}}
      \includegraphics[width=60mm]{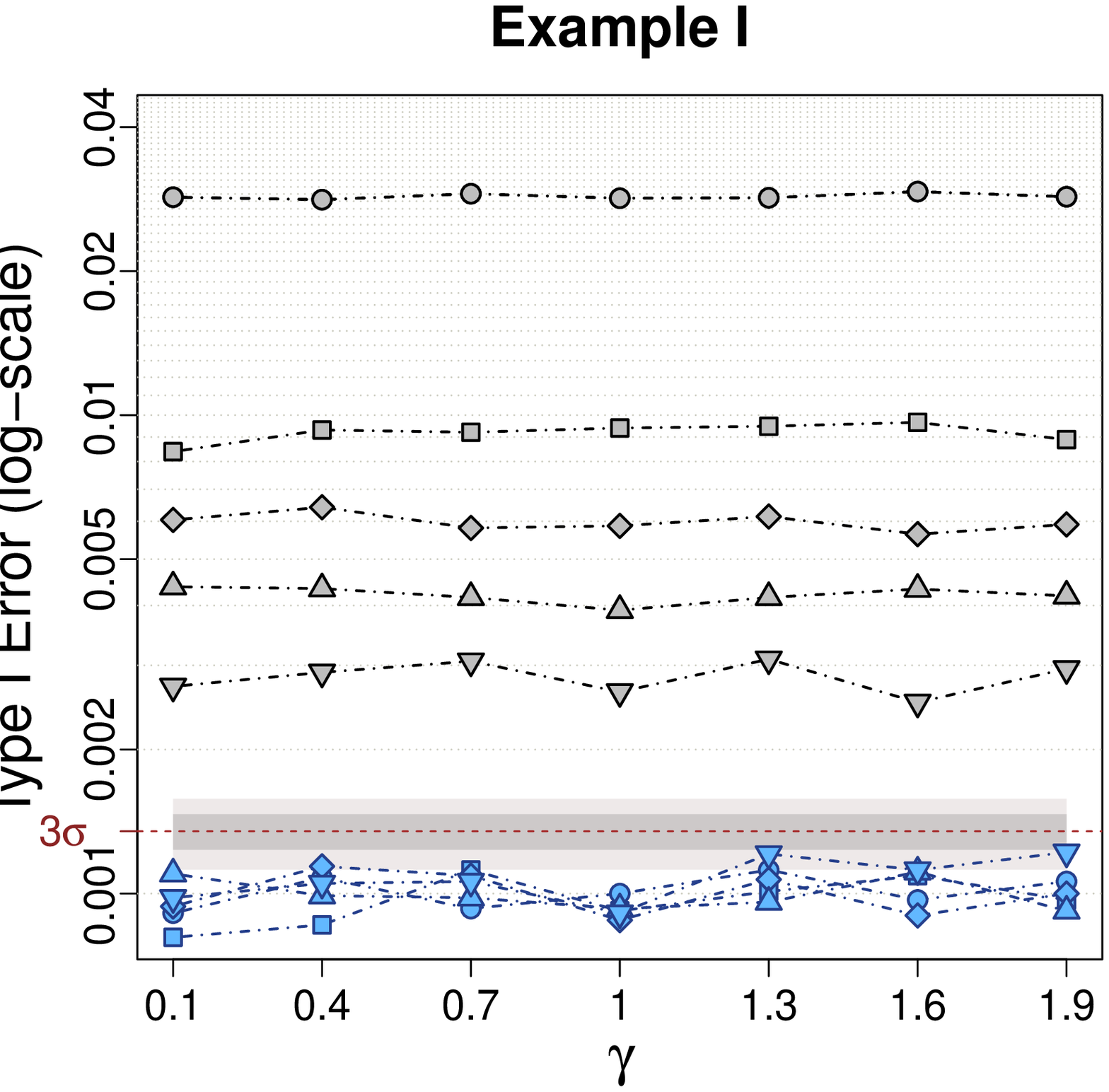}& \includegraphics[width=60mm]{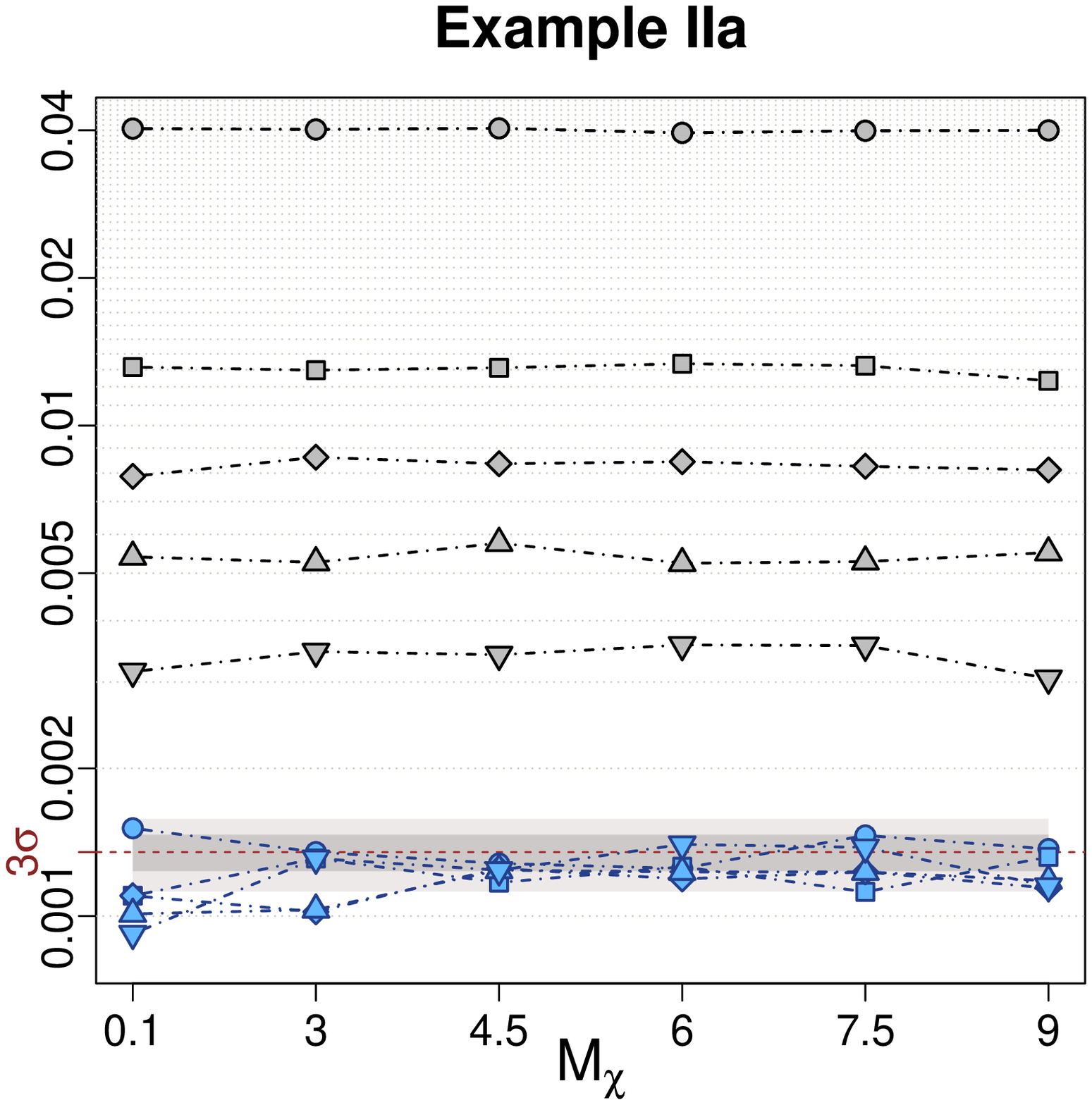}& \includegraphics[width=60mm]{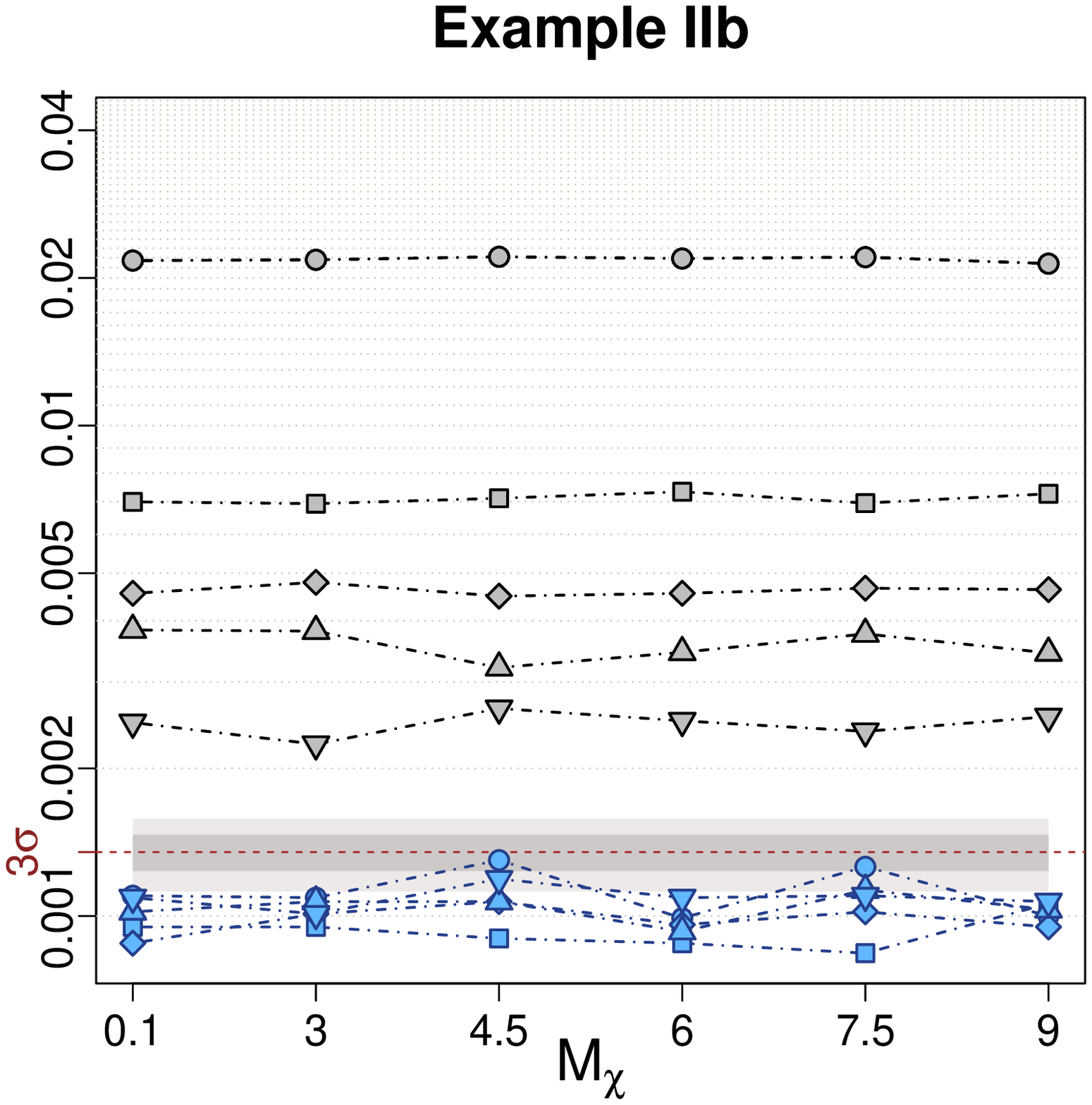} \\[-0.8cm]
      \includegraphics[width=60mm]{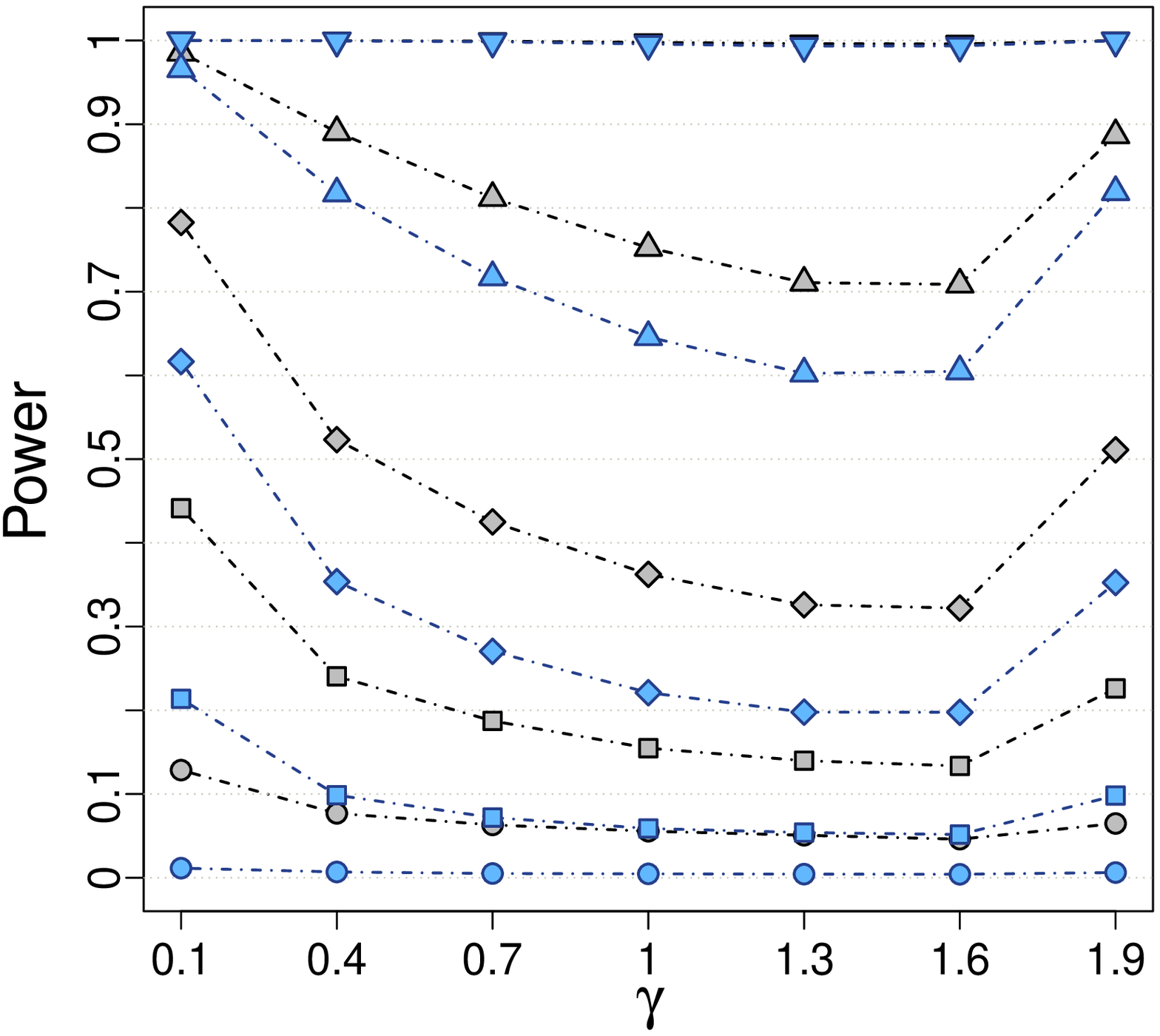} & \includegraphics[width=60mm]{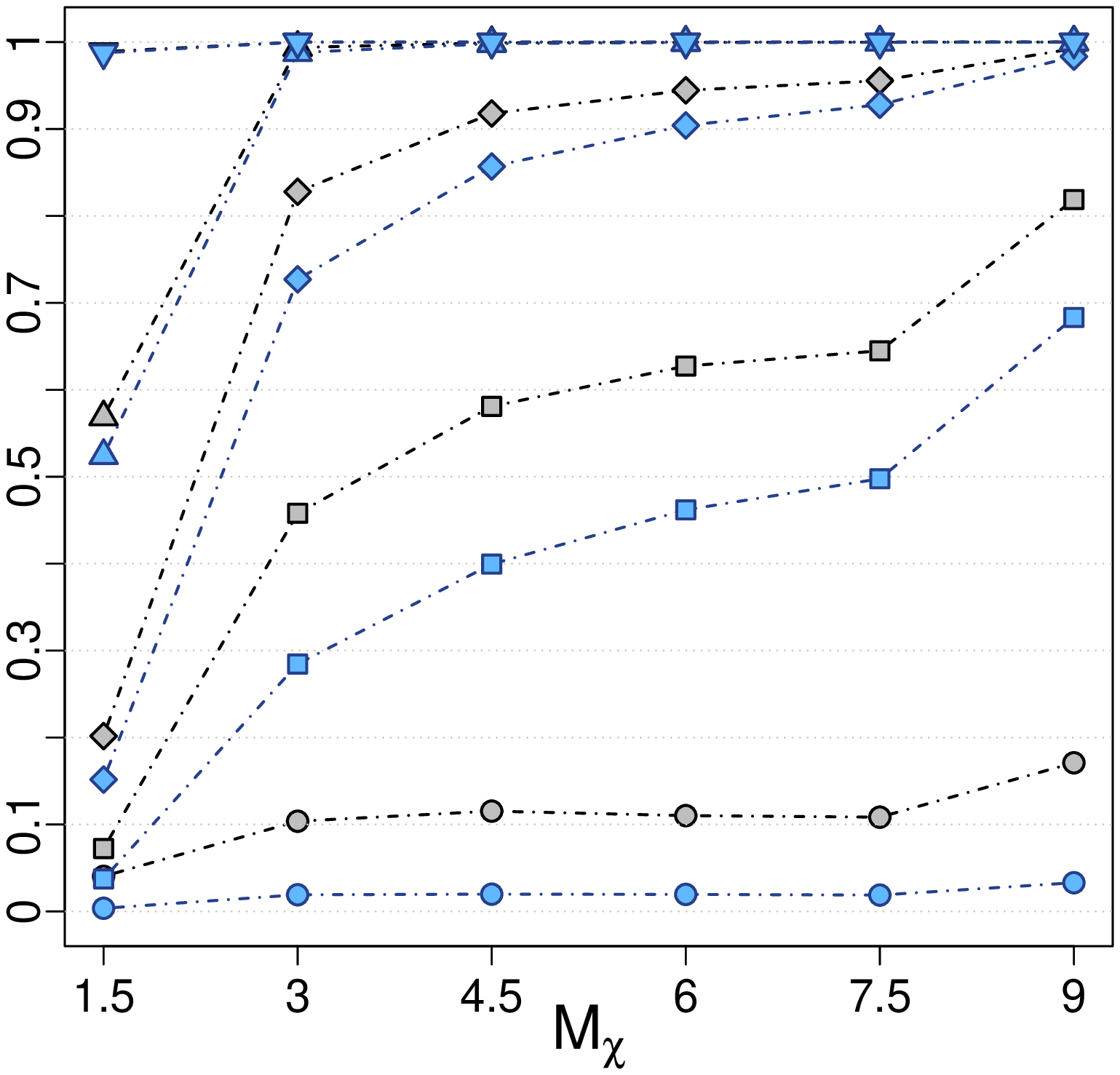}& \includegraphics[width=60mm]{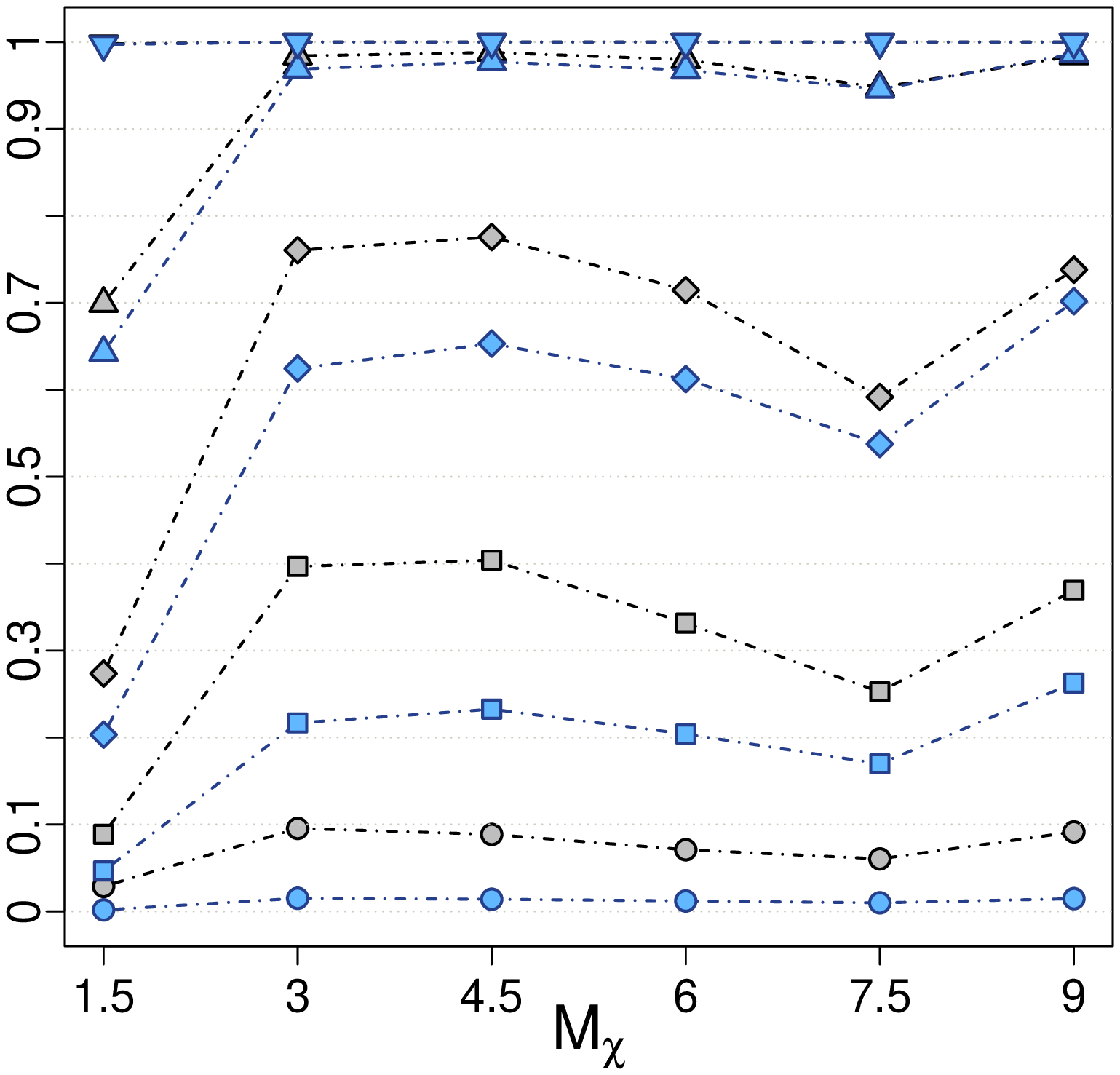} \\[-0.8cm]
      \includegraphics[width=60mm]{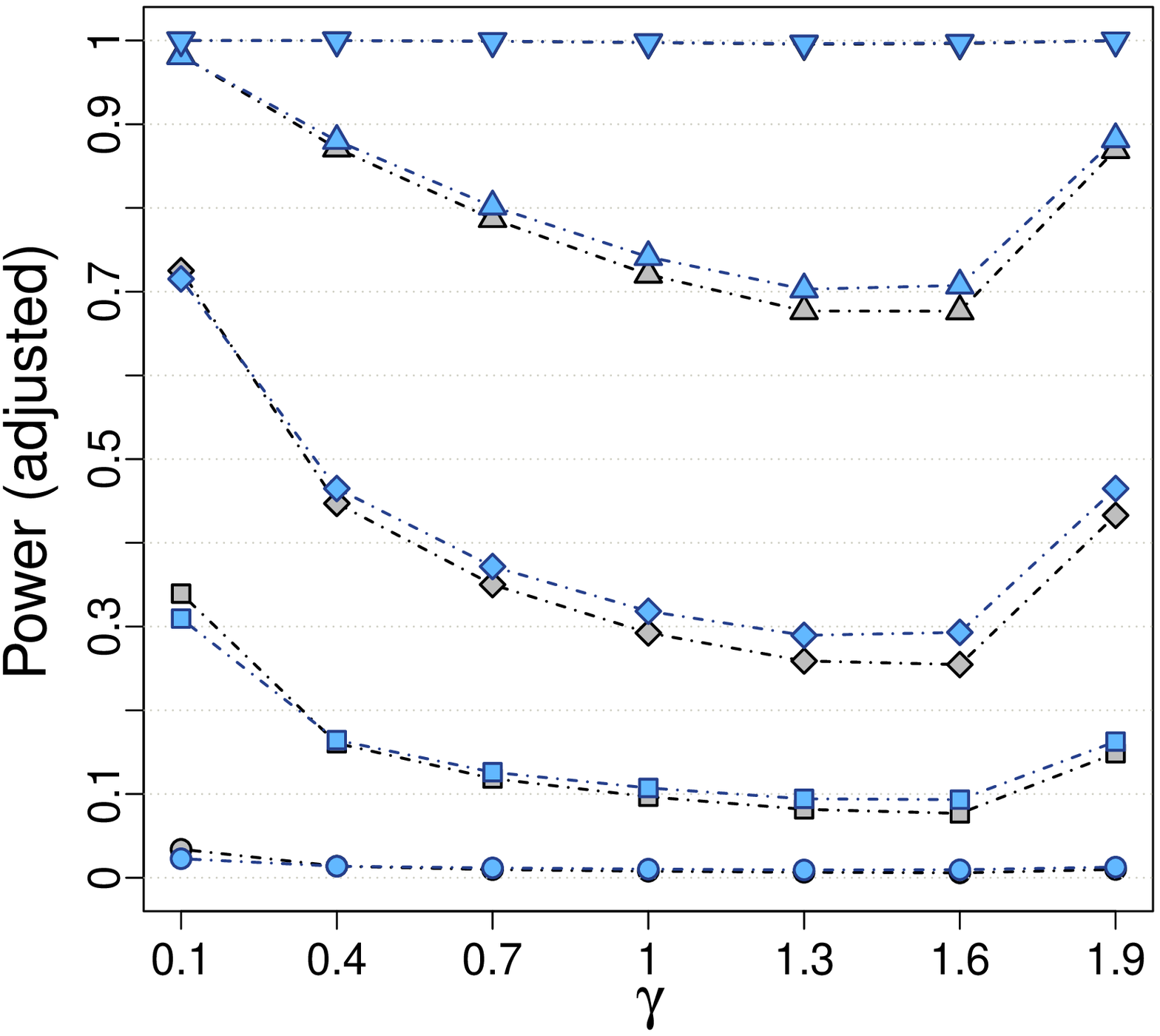} & \includegraphics[width=60mm]{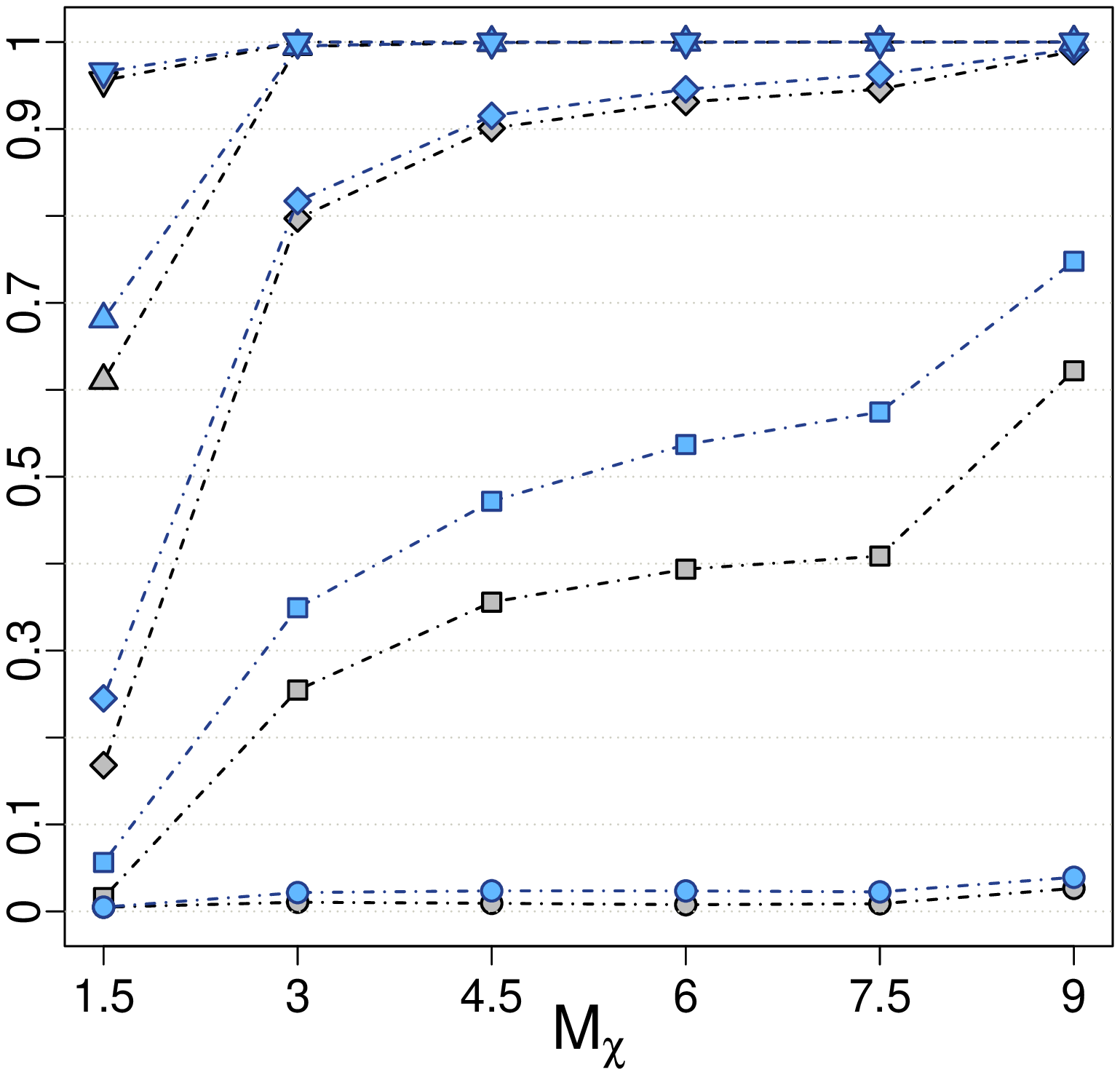}& \includegraphics[width=60mm]{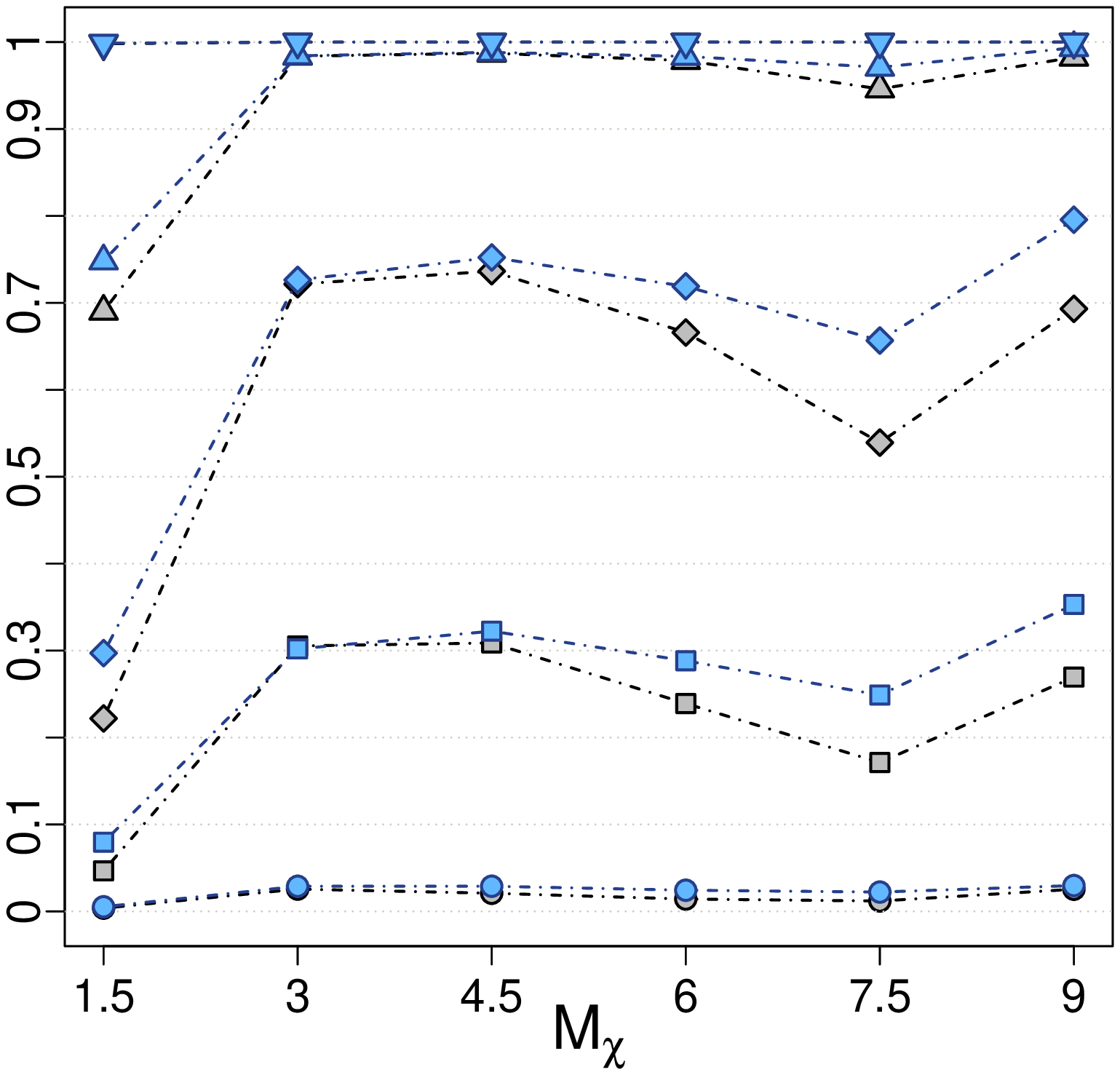} \\[-0.5cm]
\multicolumn{3}{c}{\includegraphics[width=100mm]{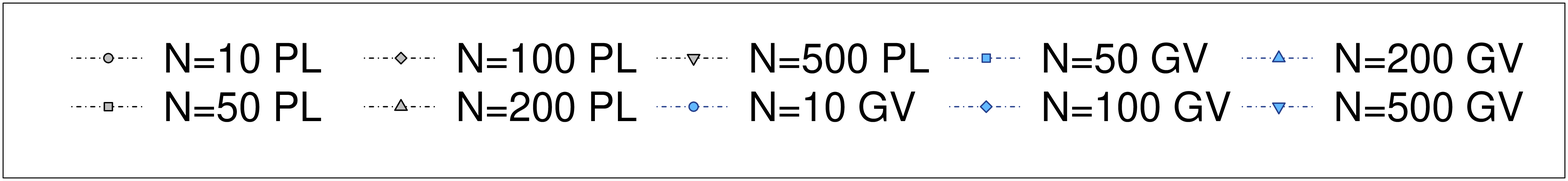}} \\
\end{tabular*}
\egroup
\hspace*{-2cm}
\caption[Figure 1]{Simulated probability of  type I error (top row), power (middle row) and adjusted power (bottom row) for  Example~I (first column), Example~IIa (second column) and Example~IIb (third column) with different sample size $N$ over $100,000$ simulations. The gray symbols corresponds to PL and the blue symbols to GV.  Shaded areas indicate regions expected to contain 68\% (dark gray) and  95\% (light gray) of the symbols if the nominal type I error of 0.0013 holds. }
\label{plots}
\end{figure*}

Results of the simulation studies appear in Fig.~\ref{plots}. Its  columns correspond to Example~I, Example~IIa and Example~IIb, respectively. In the first row, we report the simulated detection rates;  the simulated test statistics $C_\mathrm{PL}$ and $C_\mathrm{GV}$ (where $\theta$ is either $\gamma$ or $M_{\chi}$) were calculated for each of $100,000$ datasets generated from the null model. These values were then compared to the nominal  thresholds at $3\sigma$, obtained, as in (\ref{qPL}) and (\ref{qGV}), by setting $p_\mathrm{PL}$ and $p_\mathrm{GV}$  in (\ref{PLaprox}) and (\ref{GVaprox}) equal to $1-\Phi(3)=0.0013$  and solving for $c_\mathrm{PL}$ and $c_\mathrm{GV}$ respectively, i.e.,
\begin{equation}
\label{qPL}
1-\Phi(3)=\frac{\xi_0}{2\pi}P(\chi^2_{2}\geq c_\mathrm{PL}^2) + \frac{1}{2}P(\chi^2_{1}\geq c_\mathrm{PL}^2)
\end{equation}
\begin{equation}
\label{qGV}
1-\Phi(3)=\frac{P(\chi^2_{1}>c_\mathrm{GV})}{2} +E[U(c_0)|H_0]e^{-\frac{c_\mathrm{GV}-c_0}{2}}.
\end{equation}

 In the second row of Fig.\ref{plots}, we plot the local power functions; the procedure is the same as for the simulated false detection rates except the $100,000$ datasets were generated from the alternative models with $\eta=0.2$ with different values for the location parameters $\gamma$ and $M_{\chi}$. In the third row, we evaluate an adjusted version of the local power; the simulated values of $C_\mathrm{PL}$ and $C_\mathrm{GV}$ are the same as used in the plots in the second row, but  instead of comparing them with the nominal thresholds $c_{PL}$ and $c_{GV}$, we compared them with  their empirical (bootstrap) thresholds. The empirical threshold correspond to the $0.9987$ quantiles  of the $100,000$ simulated values of $C_\mathrm{PL}$ and $C_\mathrm{GV}$ generated  for the first row of Fig.~\ref{plots}, i.e., the empirical distributions of the test statistic under $H_0$.
Looking at the first row of Fig.~\ref{plots}, the simulated false detection rates associated with GV are always  consistent with the nominal $3\sigma$ error rate. This is not the case for PL. Although the false detection curves appear to approach the desired value as the sample size increases, they  are always higher than expected. Looking at the second row of Fig.~\ref{plots}, on the other hand, the simulated local power of PL is always higher than that of GV, at least the for the smaller samples sizes.  The difference between the local power functions decreases when the sample size increases, leading to two identical curves at 500 counts.
These results are, however, not sufficient to determine weather PL or GV is better. In particular, we recall our definition of  \emph{good} test as a test of hypothesis which makes the power as high as possible while keeping the false detection rate less than or equal to $\alpha_\mathrm G$, which in our examples is set to 0.0013. In this sense, the increased power of PL  is artificial; it is due to an increase of the probability of a type I error, and thus does not satisfy our \emph{goodness} requirements. Conversely, GV seems to fit in our definition of a good test of hypothesis: the false detection rate is equal to or smaller than expected, and its local power function approaches that of PL as the sample size increases. As specified in \eqref{PLaprox}, $p_{\mathrm{PL}}$ is a valid approximation to $P(C_\mathrm{PL}(\hat{\theta})>c_\mathrm{PL})$ asymptotically, i.e., for large values of $c_\mathrm{PL}$. The higher than expected type I error rate of PL in our simulations, however, does not appear to be the result of $c_\mathrm{PL}$ being too small. As described in \cite{PL05}, the error rate of $p_\mathrm{PL}$ as an approximation to $P(C_\mathrm{PL}(\hat{\theta})>c_\mathrm{PL})$ is in the order of $o(c^{-1}e^{-c^2/2})$. In our three examples the values for $c_\mathrm{PL}$ solving (\ref{qPL}) are 3.896, 3.939 and 3.937 respectively, leading to an approximation error of the order of $10^{-4}$. Thus, the high false detection rate of PL is unlikely to be due to an underestimation of the $3\sigma$ nominal thresholds. Instead, it indicates that even a sample size of 500 is not  sufficiently large to guarantee the convergence of $C_{\mathrm PL}$ to the supremum of a mean zero Gaussian process, as discussed in Section \ref{methods}. This, however, does not invalidate the utility of PL for large sample sizes as shown in \cite{PL05,PL06}.

A more detailed comparison of the detection power of PL and GV can be done by correcting the false detection rate (as in the third row of Fig.~\ref{plots}). Specifically, we can use the empirical detection threshold when evaluating the local power of the two procedures. This guarantees a false detection rate of 0.0013 ($3\sigma$ significance). GV has a lower chance of Type I error than the adjusted PL, i.e., the adjusted PL has probability 0.0013 of Type I error, which bounds that of GV, see first row of Fig.~\ref{plots}. Despite this, for all three examples and for all signal locations (values of $\gamma$ or $M_{\chi}$) considered, GV is equally or more powerful than PL when using the empirical threshold. Thus, the evidence from this simulation indicates that for small sample sizes, GV is uniformly locally more powerful than PL.

Comparing the local power functions in the second and third rows of Fig.~\ref{plots} with the pdfs in Fig.~\ref{densities}, we  see that, for Example~I, the detection power of the  testing procedures is affected by both the specific location of the signal and its spread over the search region. The power is higher when the resonance is narrowly dispersed and is located in a region with low background. In Example~II, only the location of the source emission seems to affect the power. In particular, detection appears to be more difficult in high background areas of the spectrum, and thus the strength of the signal is weaker with respect to the background sources. These issues are overcome  if at least 500 counts are available; in this case both procedure exhibit maximum detection power regardless the location or dispersion of the signal.

Few computational difficulties arose when implementing PL and GV. For PL, the most problematic step is the calculation of the geometric constant $\xi_0$ in (\ref{PLaprox}), which is computed via \eqref{xi1} for Example~I and via \eqref{xi2} for Example~II. 
This involves the numerical computation of nested integrals and it  can significantly slow down the testing procedure for complicated models. In the case of Examples I and II,  small ranges over the energy spectra $\mathcal{Y}$ ($[0;2]$ and $[1;10]$ respectively) were  chosen in order to speed up the computation of these integrals, which tended to diverge numerically over larger energy bands.  In presence of  nuisance parameters under the null model, such as $\tau$ in Example~II, the calculation of $\xi_0$ required by \eqref{xi2} is particularly complicated and considerably slower than that required by \eqref{xi1}. 

The main difficulty with GV is associated with Step 2 of Algorithm 1 in Section~\ref{methods}, which involves a multidimensional constrained optimization that must be repeated  $M$ times over  a grid, $\Theta_\mathrm{G}$, of size $R$. In Example II for instance, where $R$ is set to 50, at each of the $M=100,000$ Monte Carlo simulations, 50 two-dimensional constrained optimizations are implemented simoultaneously. If the nuisance parameter under $H_1$, $\theta$, is one-dimensional, the necessary computation can easily be accomplished by choosing a small threshold $c_0$ as described in Section~\ref{methods} and in more detail in \cite{gv10}. 
Unfortunately, using GV is more complicated when $\theta$ is multidimensional. A possible solution is proposed in \cite{vg11} in which, the number of upcrossings of the LRT process is replaced by the concept of Euler characteristics, which unfortunately does not enjoy the advantages available with the $c_0$ threshold. As discussed by the authors, the higher the number of dimensions, the higher the chances the $\chi^2$ approximation may fail as the number of regions with weak background increases. Further, increasing the dimensions, the computational effort for each Monte Carlo simulation increases drastically. Larger sample sizes are needed for each simulation in order to guarantee $\chi^2$ distribution. This, combined with the Monte Carlo simulation size needed for adequate accuracy, may lead to impractical CPU requirement. In this scenario, provided there is sufficient data to ensure an appropriate type I error rate, the numerical integrations required by PL may   be preferable. Some examples of multidimensional case are discussed in both \cite{PL05,PL06};   specifically, in \cite{PL05}, the analysis in our Example I is  further extended to a two dimensional search.

\begin{table}
\begin{tabular}{cccc}
\noalign{\global\arrayrulewidth0.08cm}
 \hline
 \noalign{\global\arrayrulewidth0.08pt}
  & Signal&  Signal  &    \\
 Method & Location& Strength &  Sig. \\
\noalign{\global\arrayrulewidth0.08cm}
 \hline
 \noalign{\global\arrayrulewidth0.08pt}
Unadjusted local &   35.82 & 0.042&  $5.920\sigma$\\
Bonferroni &  35.82 & 0.042&$5.152\sigma$\\ 
Gross \& Vitells  & 35.82 & 0.042 & $5.192\sigma$\\
Pilla et al. &  35.82 & 0.042$^{*}$ &$5.531\sigma$  \\
\noalign{\global\arrayrulewidth0.05cm}
 \hline
 \noalign{\global\arrayrulewidth0.05pt}
\multicolumn{4}{l}{\footnotesize $^{*}$Obtained afterwards via MLE by fixing the signal location} \\
\multicolumn{4}{l}{\footnotesize  to its PL estimate (see text).} \\

\end{tabular}
\caption{Summary of  multiple hypothesis testing, GV and PL on the Fermi LAT simulation. For the multiple hypothesis testing case, the smallest of $R=80$ (undadjusted local) p-values, Bonferroni's bound on the global p-value, along with GV and PL, are reported with their respective statistic.}
\label{table}
\end{table}

\section{Application to realistic data}
\label{application}
As a practical application, we perform the testing procedures discussed in Section~\ref{MC}  and \ref{methods} on a simulated observation of a monochromatic feature by the Fermi Large Area Telescope (LAT). The existence of such a feature within the LAT energy window would be an indication of new physics; of particular interest, it could result from the self-annihilation of a dark matter particle, and has consequently been the subject of several recent studies \cite{refB1,refB2,refB3}.
\begin{figure} 
\includegraphics[width=90mm]{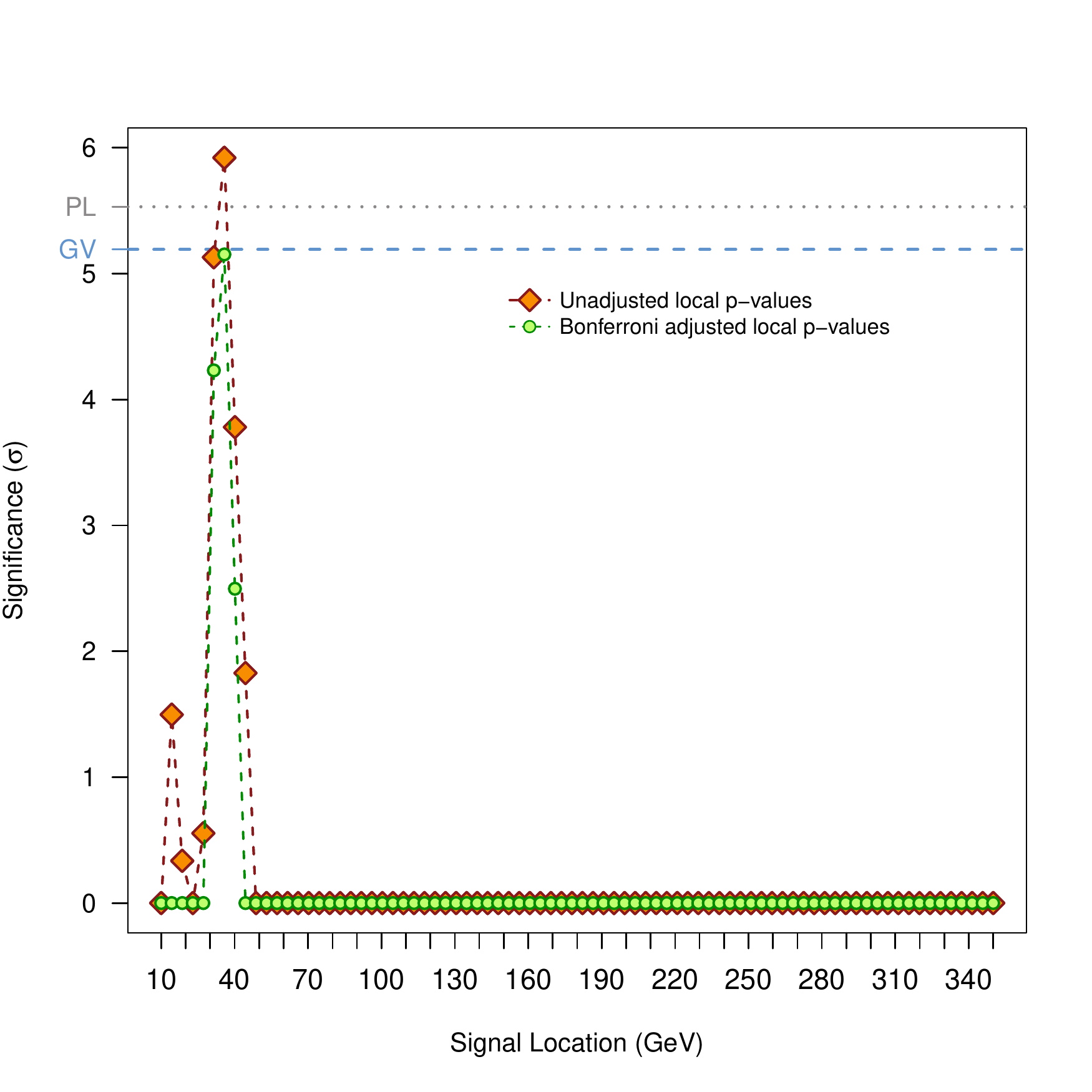} 
\caption[Figure 3]{Unadjusted local p-values (orange diamonds), Bonferroni adjusted local p-values (green dots), PL global p-value (gray dotted line) and GV global p-value (blue dashed line) for the Fermi LAT simulation. The Bonferroni's bound on the global p-value is only slightly more conservative than the GV p-value}
\label{real}
\end{figure}
We consider emission resulting from the self-annihilation of a particle making up the substantial dark matter mass of the Virgo galaxy cluster (distributed according to \cite{refB4}). We further specify that the particle have a mass of 35 GeV and a direct-to-photon thermally-averaged annihilation cross section of $1\times10^{-23}$ cm$^2$. Competing with this signal, we introduce a simple astrophysical background corresponding to isotropic emission following a spectral power-law with index $2.4$, i.e., $\tau=1.4$. Both signal and background models are then simulated for a five-year observation period using the \emph{gtobssim} package, available at \url{http://fermi.gsfc.nasa.gov/ssc/data/analysis/software}, which takes into account details of the instrument and orbit. The setup yields, on average, 64 signal and 2391 background events.

The full model is the same as in Example~II i.e., as given in (\ref{ex2}); results of the several methods are shown in Table~\ref{table} and Fig.~\ref{real}. In the multiple hypothesis testing analysis, the smallest of the local p-values is reported along with the respective estimates for the signal strength and location. As discussed in Section~\ref{methods}, the latter are  equivalent to those obtained with GV. The test statistic of PL, $C_\mathrm{PL}(\hat{\theta})$, is constructed under the assumption that $\eta=0$, and thus  does not depend on the  signal strength.  However, it does depend on the location of the source emission, and thus the estimation of $\eta$ under $H_1$ must be conducted once the signal location has been estimated (through MLE for instance). In our analysis, the PL estimate for the source location  is equivalent to both that of GV and of the local p-values methods; it follows that the resulting MLE for the signal strength is the same for all methods.

The local p-value approach leads to the largest significance of $5.920\sigma$, followed by PL $5.531\sigma$, GV $5.192\sigma$ and finally Bonferroni with $5.152\sigma$. Although PL provides the most significant of the global p-values, it is difficult to interpret this result given PL's higher than expected rate of false detections in the simulation study. The Bonferroni adjusted local p-value, over the set of 80 simultaneous tests, it is only slightly more conservative than GV. The disparity between the two is expected to grow, however,  as the number of grid points over the energy spectrum increases.

\begin{table}
 \centering
\begin{tabular}{cccc}
\noalign{\global\arrayrulewidth0.08cm}
 \hline
 \noalign{\global\arrayrulewidth0.08pt}
  & Unadjusted & Bonferroni &  Gross    \\
  & local& adj. local&  \& Vitells \\
\noalign{\global\arrayrulewidth0.08cm}
 \hline
 \noalign{\global\arrayrulewidth0.08pt}
Bkg only & 97056& 37 & 2907\\
Time (secs)& 0.974& 0.000 & 136.282\\
\hline
Bkg$+$sig & 10496& 45210 & 44294 \\
Time (secs) & 1.061& 0.000 & 137.532\\
\noalign{\global\arrayrulewidth0.05cm}
 \hline
 \noalign{\global\arrayrulewidth0.05pt}
\end{tabular}
\caption{Summary on the analysis of 100,000 simulated datasets from Example II in Section~\protect\ref{simulation}.  We report the number of times each testing method is used by the  sequential approach to make a final decision at $3\sigma$, and the respective average computational times.  The first two lines refer to the background only simulations and  whereas the last two lines correspond to the background + signal simulations. }
\label{tabcounts}
\end{table}

\section{ A  sequential approach}
\label{step}

The PL and GV methods are typically used to overcome the over-conservativeness of the  Bonferroni's bound. 
Thus, one might expect the global p-values $p_{\mathrm GV}$ and $p_{\mathrm PL}$ to be smaller or equal to $p_{BF}$. Unfortunately, this is not always true; for the specific case of GV, combining (\ref{GVaprox}) and (\ref{pL}), we have
\begin{equation}
\label{new1}
p_{\mathrm GV}= p_{\mathrm L}+E[U(c)|H_0]\leq p_L+p_{\mathrm BF}=(R+1) p_{\mathrm L}.
\end{equation}
Where $E[U(c_{\mathrm GV})|H_0]=E[U(c_0)|H_0]e^{-\frac{c_\mathrm{GV}-c_0}{2}}$ is the expected number of upcrossings of the observed value for the test statistic $c_{\mathrm GV}$, i.e., 
$c_{\mathrm GV}=LRT_{\theta^\star}$ in (\ref{pL}). Since the expected number of upcrossings above $c_{\mathrm GV}$ is bounded by the expected number of times the LRT process takes a value greather than  $c_{\mathrm GV}$, i.e., $Rp_{\mathrm L}=p_{\mathrm BF}$, and given the asymptotic equivalence of GV and PL for large sample size (see Section \ref{methods}), we have 
 \begin{equation}
\label{new2}
p_{\mathrm PL}\approx p_{\mathrm GV}\leq \frac{R+1}{R}p_{\mathrm BF}\approx p_{\mathrm BF}\quad \text{for large $R$}.
\end{equation}
For small $R$, the bound in (\ref{new2}) allows Bonferroni to provide a sharper bound than either GV or PL.
A more formal justification of \ref{new1} and \ref{new2} can be found in \cite{algeri}. 
 
\begin{figure} 
\includegraphics{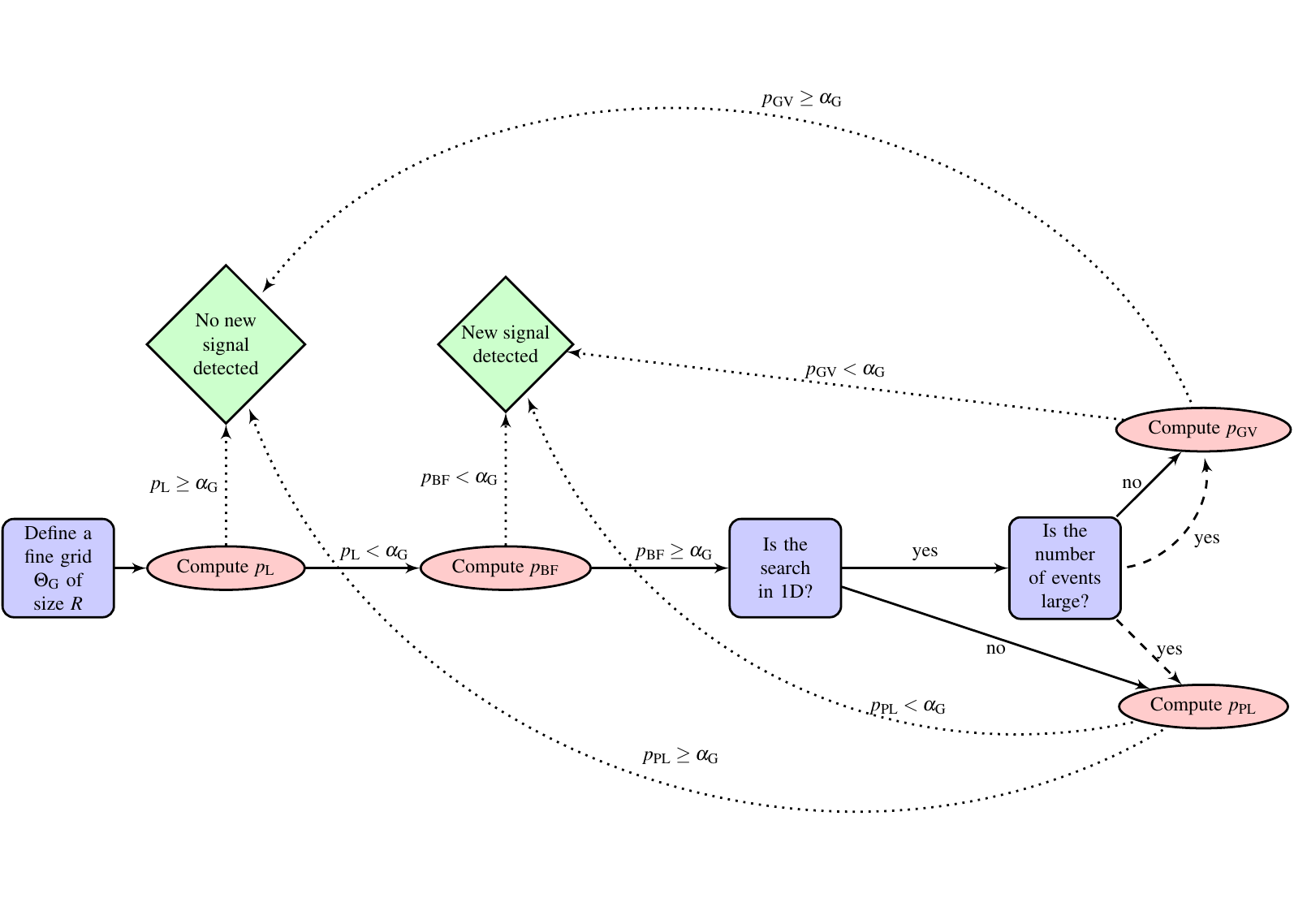} 
\caption{Outline of the sequential approach. General guidelines for statistical signal detections in HEP. $\Theta_\mathrm{G}$ is the grid of possible signal-search locations; its resolution is given by $R$. $p_\mathrm L$ is the minimum of the local p-values and $p_\mathrm{BF}$ its Bonferroni adjusted counterpart. $\alpha_\mathrm G$ is the predetermined false detection rate. $p_\mathrm{PL}$ and $p_\mathrm{GV}$ are the global p-values provided by PL \cite{PL05,PL06} and GV \cite{gv10} respectively. Dashed arrows indicate that two actions are equally valid, and dotted lines lead to the final conclusion in terms of evidence in favor of the new resonance.}
\label{diag}
\end{figure}

Based on this and the results of the previous sections, it is possible to establish  general guidelines for selecting the appropriate statistical testing procedure. 
The goal is to adhere a prescribed false-positive rate as closely as possible while minimizing  computational effort.  This can be accomplished by  combining the simplicity of multiple hypothesis testing with the robustness of global p-values in a multi-stage procedure. 
Specifically, Fig.~\ref{diag} summarizes a simple step-by-step algorithm where multiple hypothesis testing methods are implemented first, and the more time-consuming GV and PL are implemented only if simpler methods exhibit poor type I error rates and/or power.

We focus on the case of a one-dimensional search. In which,
\begin{equation}
\label{order1}
p_{\mathrm L}\leq  p_{\mathrm PL}\approx p_{\mathrm GV}\lessapprox p_{\mathrm BF},
\end{equation}
where the approximation sign in the last inequality allows the situation discussed above where $p_{\mathrm GV}\geq p_{\mathrm BF}$ . Despite this possibility, the bound in (\ref{order1}) is an approximation for large $R$, where $\frac{R+1}{R} \approx 1$. 

In order to implement the sequential approach, we first calculate the $R$ unadjusted local p-values over the grid $\Theta_{\mathrm G}$; the minimum of these p-values is denoted by $p_{\mathrm L}$. From (\ref{order1}), if we observe $p_{\mathrm L}>\alpha_{\mathrm G}$ we fail to reject reject $H_0$ with any of the procedures and we can immediately conclude that we cannot reject $H_0$.  On the other hand, if $p_{\mathrm L}\leq \alpha_{\mathrm G}$, a correction for the simultaneous $R$ tests is needed, and because of its easy implementation, we compute $p_{\mathrm BF}$. Whereas, if $p_{\mathrm BF}< \alpha_{\mathrm G}$, then all methods reject $H_0$, and we can claim evidence in favor of the new source. Conversely, if $p_{\mathrm BF}\geq \alpha_{\mathrm G}$ we should implement a method that is typically less conservative than Bonferroni's correction, when dealing with large significances (e.g. $3\sigma, 4\sigma, 5\sigma$), such as GV or PL. Specifically, on the basis of the simulations in Section~\ref{simulation}, GV appears to be preferable for small sample sizes, as it provides a  false-positive rate less than or equal to $\alpha_{\mathrm G}$. For large sample sizes, PL and GV are equivalent, and the decision between GV and PL depends on the details of the models compared. As discussed in Section~\ref{simulation}, PL requires extensive numerical integration which can diverge for large search windows $\Theta$, while GV requires a small number of Monte Carlo simulations which might become troublesome for complicated models.  Finally, if $p_{\mathrm GV}< \alpha_{\mathrm G}$ (or $p_{\mathrm PL}< \alpha_{\mathrm G}$) we can claim  evidence in support of the new resonance, whereas if  $p_{\mathrm GV}\geq \alpha_{\mathrm G}$ (or $p_{\mathrm PL}\geq \alpha_{\mathrm G}$) we cannot claim that a signal has been detected.

\begin{table}
 \centering
\begin{tabular}{ccc}
\noalign{\global\arrayrulewidth0.08cm}
 \hline
 \noalign{\global\arrayrulewidth0.08pt}
& Type I error & Power  \\
\noalign{\global\arrayrulewidth0.08cm}
 \hline
 \noalign{\global\arrayrulewidth0.08pt}
  Unadjusted local  & 0.03033& 0.89502\\
  Bonferroni adj. local  &0.00040& 0.45211\\
Gross \& Vitells  & 0.00089& 0.53159 \\
 Sequential approach  & 0.00087& 0.53161 \\
\noalign{\global\arrayrulewidth0.05cm}
 \hline
 \noalign{\global\arrayrulewidth0.05pt}
\end{tabular}
\caption{Probability of type I error and power of the testing methods and sequential approach implemented on 100,000 simulated datasets from Example II in Section~\protect\ref{simulation}. }
\label{tabproperties}
\end{table}

The sequential approach  involves choosing a procedure based on the characteristics of the data. Thus, one might be concerned about possible ``flip-flopping'' similar to that described by Feldman and Cousins in \cite{flipflop} in the context of confidence intervals. As argued below, however, this is not the case for the sequential approach illustrated in Fig.~\ref{diag}.
By virtue of (\ref{order1}), both the type I error and the power of the   sequential approach are approximately equivalent to those of GV (or PL) for large values of $R$. For clarity, we hereinafter suppose GV is used rather than PL in the   sequential approach. The statistical results follow in exactly the same way however, if PL is used for large sample sizes. 

Let $\tilde{\alpha}$ be the false detection rate associated with the  sequential approach, and consider the events
\begin{equation*}
\begin{split}
{\mathrm BF}_0&=\{\text{Reject $H_0$ at level  $\alpha_{\mathrm G}$ with Bonferroni}\}\\
{\mathrm GV}_0&=\{\text{Reject $H_0$ at level  $\alpha_{\mathrm G}$ with GV}\}.\\
\end{split}
\end{equation*}
As in (\ref{alphapower}) we use $P(\cdot|\eta=0)$ to denote the probability that one event occurs given that the null hypothesis is true, i.e., in absence of the signal. Because the sequential approach rejects $H_0$ when either Bonferroni or GV does so, it follows that

{\small
\begin{equation*}
\begin{split}
\tilde{\alpha}&=P({\mathrm BF}_0 \text{ or } {\mathrm GV}_0|\eta=0)\\
&=P({\mathrm BF}_0|\eta=0 )+P( {\mathrm GV}_0|\eta=0)-P({\mathrm BF}_0  \text{ and }  {\mathrm GV}_0|\eta=0)\\
&=P({\mathrm BF}_0|\eta=0 )+P( {\mathrm GV}_0|\eta=0)-P({\mathrm GV}_0|{\mathrm BF}_0,\eta=0)P({\mathrm BF}_0|\eta=0).\\
\end{split}
\end{equation*}}
By the ordering of the p-values in (\ref{order1}), if $H_0$ is rejected by Bonferroni, then it is typically rejected by GV and thus, \[P({\mathrm GV}_0|{\mathrm BF}_0,\eta=0)\approx1,\] from which it follows that
$\tilde{\alpha}\approx P( {\mathrm GV}_0|\eta=0)$, where $P( {\mathrm GV}_0|\eta=0)$ is the false detection rate of GV.
The power of the    sequential approach can be obtained in a similar manner by considering the events 
\begin{equation*}
\begin{split}
{\mathrm L}_1&=\{\text{Reject $H_0$ at level  $\alpha_{\mathrm G}$ with local p-values}\}\\
{\mathrm GV}_1&=\{\text{Reject $H_0$ at level  $\alpha_{\mathrm G}$ with GV}\},\\
\end{split}
\end{equation*}
and evaluating probabilities of the type $P(\cdot|\eta, \theta)$ defined in (\ref{alphapower}).

\begin{figure}
\includegraphics{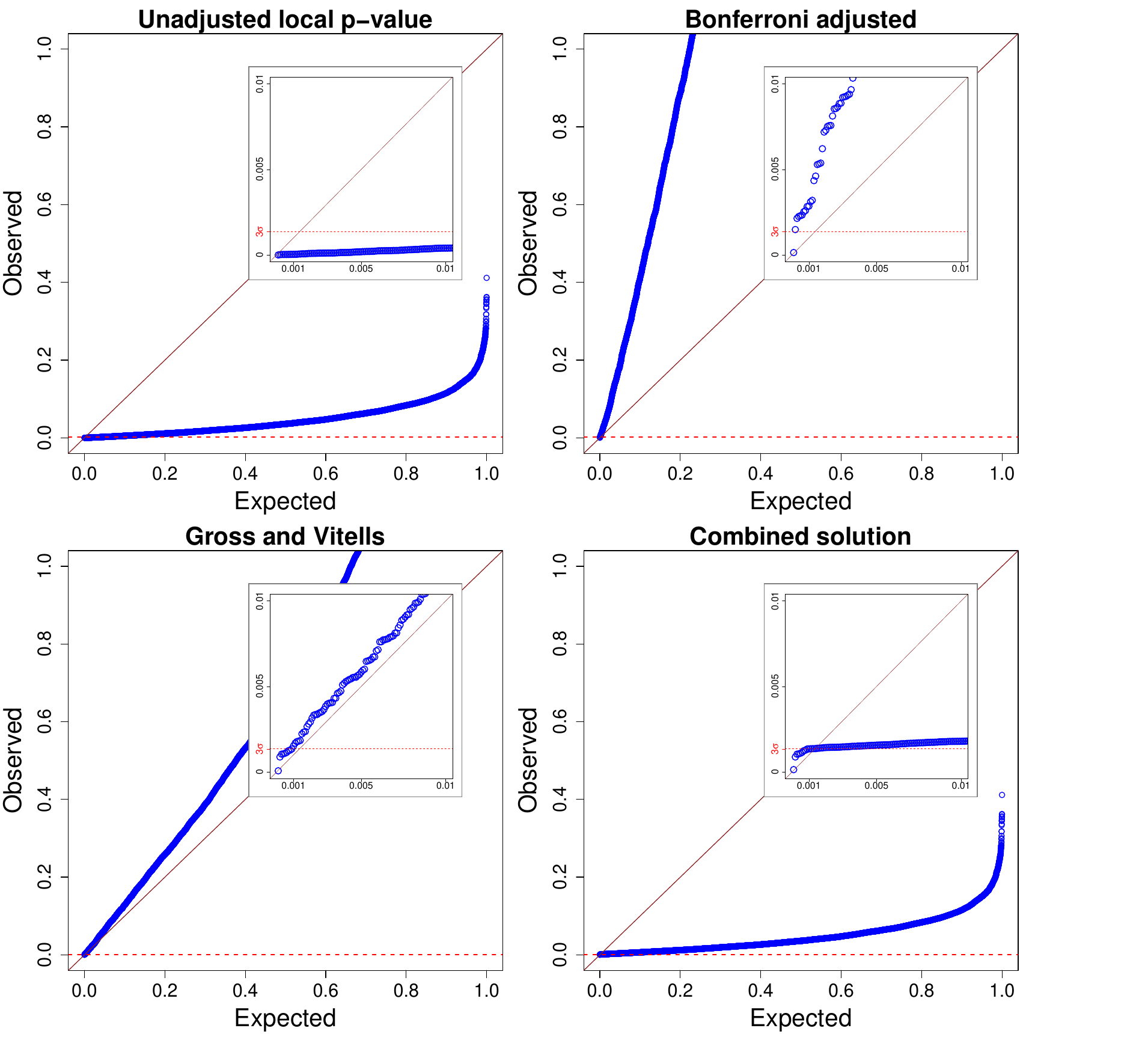} 
\caption{QQ-plots for the unadjusted local, Bonferroni's bound and GV p-values computed for the 100,000 simulated background-only datasets from Example II of Section~\protect\ref{simulation}. Each dataset considers 2000 background only events. The p-values selected via the    sequential  procedure in Fig.~\protect\ref{diag} are also reported. Each set of p-values is compared with the expected quantiles of a Uniform distribution on $[0,1]$.  The inlayed plots in each panel magnify the important range of the p-value distributions near zero.}
\label{qqplots}
\end{figure}

To illustrate its statistical properties, we apply the combined approach to a set of 100,000 simulated datasets from the model in Example II with $\tau$ fixed at 1.4.  For each dataset we first simulate 2000 background only events and then we simulate 30 additional events from a Gaussian source centered at 9 GeV. For both the 100,000 background only datasets and the 100,000 background plus source datasets we compute unadjusted local p-values, Bonferroni's corrections,  and  GV.
Table~\ref{tabcounts} reports the number of times each of the testing procedures considered is selected by the  sequential approach to make a final decision at the $3\sigma$ significance level. The average computational times for each method are also reported. In the presence of source emission, the most computationally expensive method GV was used only about $44\%$ of the time, leading to a computational gain of about 89 days over the 100,000 simulations. Conversely, in absence of the signal,  GV was used about $2.9\%$
of the time, leading to a computational gain of about 155 days. In order to assess the robustness of the method with respect to the desired statistical properties, we computed the false discovery rate and the power using nominal levels at $3\sigma$ significance. The results are presented in Table~\ref{tabproperties}. As discussed above, the  sequential approach  exhibits  statistical properties which are approximately equivalent to those of  GV (or PL). As expected, the small discrepancies between the two methods are due to the fact that in $0.375\%$ of the replications $p_{\mathrm GV}>p_{\mathrm BF}$. When removing these cases from the analysis, both the probability of a Type I error and the power of the  sequential approach coincide with those of GV.

Finally, Fig.~\ref{qqplots} displays the p-values computed with each procedure on each of the 100,000 simulated background-only datasets. Ideally a p-value will follow a uniform distribution on the unit interval under repeated sampling of data under $H_0$: this insures that the method will have the target Type I error rate. In the QQ-plots in Fig.~\ref{qqplots}, the p-values will fall along the $45^\circ$ line if they follow a uniform distribution. If they deviate above this line, the procedure is conservative and if they deviate below the procedure will exhibit too many false positives. As expected, the unadjusted local p-values are always smaller than their expected values assuming uniform distribution, whereas both Bonferroni and GV are  conservative. The sequential approach leads to an intermediate situation in which the p-values are over-conservative up to the significance level $\alpha_G$ adopted at each step  of the algorithm in Fig.~\ref{diag} ($3\sigma$ in Fig.~\ref{qqplots}), whereas the p-values become under-conservative above  $\alpha_G$, i.e., only for uninteresting cases.

\section{Discussion}
\label{discussion}
In this article we  investigate  the performance of four different testing procedures for the statistical detection of new particles: the multiple hypothesis testing approach based on local p-values \cite{dellanegra, dvd14}, its Bonferroni adjusted counterpart, the LRT-based approach of Gross and Vitells \cite{gv10}, GV, and the Score-based approach of Pilla et al. \cite{PL05,PL06}, PL. To the best of our knowledge, ours is the first application  in a realistic scientific problem of PL in \cite{PL06}, i.e., in presence of nuisance parameters under $H_0$.

We  show  analytically that local p-values are strongly affected by the arbitrary choice of the grid resolution, $R$, over the energy range where the tests are conducted. Specifically, when $R$ is sufficiently large, the unadjusted p-values provide a higher number of false detections than expected, whereas the Bonferroni's bound on the global p-value may lead to over conservative inference  if $R$ is large. However, as shown in our realistic data analysis, if $R$ is only moderately large
($R=80$ in our case) Bonferroni represents a reasonable choice. Additionally, cases may arise where Bonferroni's bound leads to less stringent acceptance criteria than GV and PL. Thus, in order to make final conclusions and to take advantage of the easy implementation of the Bonferroni correction, it should always  be used as a preliminary tool in statistical signal detection as described in Section~\ref{step}. 

If the number of search regions $R$ is quite large, a good trade-off is provided by both PL and GV which produce global p-values as a measure of the evidence for a new source of emission. Although, PL and GV lead to the same conclusions for large sample sizes, based on our simulations,  for small samples sizes PL may produce a higher number of false detections than expected. This strongly compromises the reliability of PL when only a few events are available,  and thus GV is preferable in this case.
From a computational perspective, difficulties may arise with both methods when dealing with complex models; these stem  from the required numerical integrations of PL and the Monte Carlo simulations and multidimensional optimization of GV. The latter are not required by PL since the procedure does not require estimation of the signal strength.

PL  requires  a higher level of mathematical complexity to compute the geometric constants involved. This is exacerbated  when free parameters are present under the null model, and the methodology must be extended as in \cite{PL06}.  On the other hand, PL can automatically be implemented when the  nuisance parameter under the alternative hypothesis is multidimensional, whereas the existing multivariate counterpart of GV \cite{vg11} relies on the computation of Euler characteristics, which does not enjoy the simplicity and computational efficiency of the one-dimensional case. 

Section~\ref{step} summarizes the methods and provides step-by-step guidelines for a  sequential approach for statistical signal detection in High Energy Physics.  The   sequential approach preserves both false detection rate and power, while allowing  considerable gains in terms of implementation and computational time relative to other methods.

\section{Acknowledgement}
JC thanks the support of the Knut and Alice Wallenberg foundation and the Swedish Research Council. DvD acknowledges support from a Wolfson Research Merit Award (WM110023) provided by the British Royal Society and from Marie-Curie Career Integration (FP7-PEOPLE-2012-CIG-321865) and Marie-Skodowska-Curie RISE (H2020-MSCA-RISE-2015-691164) Grants both provided by the European Commission.

\begin{appendix}
\section{Appendix}
\subsection{Covariance function of $\{C^\star_{\mathrm PL}(\theta),\theta \in \Theta\}$}
\label{appendix1}
If the nuisance parameter under $H_0$, $\mathbf{\phi}$, is known, the covariance  function $W(\theta,\theta^{\dag})$ in \eqref{Sstat} of $\{C^\star_{\mathrm PL}(\theta),\theta \in \Theta\}$ is given by
\begin{equation}
\label{W1}
W(\theta,\theta^{\dag})=\int_\Theta\frac{g(y,\theta)g(y,\theta^\dag)}{f(y,{\bm, \phi})}d\theta-1.
\end{equation}
Conversely, if $\mathbf{\phi}$ is unknown,  it  is replaced by its MLE under $H_0$ in \eqref{CPLstar} and the covariance function $W(\theta,\theta^{\dag})$ is modified accordingly. For illustration, we consider the case where $\phi$ is one-dimensional and
$W(\theta,\theta^{\dag})$ is given by
\begin{equation}
\label{W2}
W(\theta,\theta^{\dag})=W_\mathbf{\phi}(\theta,\theta^{\dag})-\frac{W(\theta|\hat{\phi}_0)W(\theta^\dag|\hat{\phi}_0)}{I(\hat{\phi}_0)},
\end{equation}
where $\hat{\phi}_0$ is the MLE of $\phi$ under $H_0$, $I(\hat{\phi}_0)$ is the Fisher information $\frac{\partial^2 \log f(y,\phi)}{\partial^2 \phi}$ under $H_0$ evaluated at $\hat{\phi}_0$, and $W(\theta|\hat{\phi}_0)=\bigintss g(y,\theta)\frac{\partial \log f(y,\phi)}{\partial \phi}|_{\phi=\hat{\phi}_0}d y$. The  multi-dimensional generalization of \eqref{W2} is described in  \cite{PL06}.

\subsection{Geometric constant $\xi_0$ in the calculation of $p_\mathrm{PL}$}
\label{appendix2}
If the nuisance parameter under $H_0$, $\mathbf{\phi}$, is known, the geometric constant $\xi_0$ in \eqref{PLaprox} is given by
\begin{equation}
\label{xi1}
\xi_0=\bigint_{\Theta}\frac{\sqrt{\biggl| W(\theta,\theta^{\dag})\frac{\partial^2 W(\theta,\theta^{\dag})}{\partial \theta \partial \theta^\dag}-\frac{\partial W(\theta,\theta^{\dag})}{\partial \theta }\frac{\partial W(\theta,\theta^{\dag})}{\partial \theta^\dag} \biggl|_{\theta^\dag=\theta}}}{W(\theta,\theta)} d\theta.
\end{equation}
Whereas, if $\mathbf{\phi}$ is unknown, $\xi_0$ is given by
\begin{equation}
\label{xi2}
\xi_0=\bigintss_{\Theta}\sqrt{\frac{\partial^2 \rho^\star(\theta,\theta^\dag)}{\partial \theta \partial \theta^\dag}} \biggl|_{\theta^\dag=\theta} d\theta\qquad \text{with $\rho^\star(\theta,\theta^\dag)=\frac{W(\theta,\theta^{\dag})}{\sqrt{W(\theta,\theta)W(\theta^{\dag},\theta^{\dag})}}$}.
\end{equation}
Given the complexity of \eqref{xi1} and \eqref{xi2}, their computation typically required numeric integration.

\end{appendix}

\end{document}